\RequirePackage{fix-cm}
\documentclass[smallextended]{svjour3}       
\usepackage{amsmath}
\smartqed  
\usepackage{graphicx}
\usepackage{multirow}
\usepackage{epsfig}
\usepackage{epstopdf}
\usepackage{amsfonts}

\journalname{}

\title{MUBs and SIC-POVMs of a spin-1 system from the Majorana approach}

\author{P.K. Aravind }

\authorrunning{P.K. Aravind}

\institute{P.K.Aravind$^{1}$ \at
$^{1}$Physics Department, Worcester Polytechnic Institute, Worcester, MA 01609, U.S.A.\\
\email{paravind@wpi.edu}}

\date{\today}

\begin{document}
\maketitle
\begin{abstract}

In the Majorana or stellar representation of quantum states, an arbitrary (pure) state of a spin-1 system is represented by a pair of points on the unit sphere or, equivalently, by a pair of unit vectors. This paper presents an expression for the squared modulus of the inner product of two spin-1 states in terms of their Majorana vectors and uses it to give a geometrical construction of the MUBs and SIC-POVMs of a spin-1 system. The results are not new and duplicate those obtained earlier by other methods, but the Majorana approach nevertheless illuminates them from an  unusual point of view. In particular, it reveals the MUBs and SICs as symmetrical collections of vectors in ordinary three-dimensional space, rather than as rays in a projective Hilbert space. While it does not appear feasible to extend this treatment to higher spin systems, the spin-1 case exhibits sufficient subtlety and complexity to be worth spelling out for its pedagogical and historical interest.

\end{abstract}

\section{\label{sec:Intro}Introduction}

Mutually unbiased bases (MUBs) \cite{ivanovic}-\cite{weiner} and Symmetric Informationally Complete Positive Operator Valued Measures (SIC-POVMs, or SICs for short) \cite{zauner}-\cite{grassl2017} are important features of finite quantum systems that have been studied both for their interest in connection with the foundations of quantum mechanics and their practical applications \cite{bennett}-\cite{mandayam2}. We first recall their definitions in a $d$-dimensional Hilbert space.\\

Two orthonormal bases, $|\psi^{(1)}_{i}\rangle$ and $|\psi^{(2)}_{i}\rangle$ for $1 \leq i \leq d$, are said to be mutually unbiased if $|\langle\psi^{(1)}_{i}| \psi^{(2)}_{j}\rangle|^{2}=\frac{1}{d}$ for all $i,j = 1,...,d$. A set of bases is said to be mutually unbiased if every pair among them is mutually unbiased. The maximum number of MUBs in a Hilbert space of dimension $d$ is $d+1$ and, if $d$ is a prime or a prime power, explicit constructions of such sets are known \cite{ivanovic}-\cite{durt}.\\

Turning next to SICs, a set of $d^{2}$ normalized states $|\psi_{j}\rangle$ for $j=1,...,d^{2}$ is said to form a SIC if $|\langle\psi_{i}| \psi_{j}\rangle|^{2}=\frac{1}{d+1}$ for all $i\neq j$. It was conjectured by Zauner \cite{zauner} that SICs exist in all finite dimensions. Numerous examples of SICs in many dimension are known and several recent articles \cite{Appleby2017a}-\cite{grassl2017} have used numerical or other approaches to push the dimensions in which SICs are known into the low hundreds, with one paper \cite{grassl2017} actually finding a solution in $d = 844$. However a formal proof of Zauner's conjecture is still lacking. Both MUBs and SICs are of interest in connection with quantum tomography, or the problem of determining an unknown quantum state of which a number of copies are available \cite{ivanovic},\cite{fields},\cite{fuchs}; MUBs use only projective (or von Neumann) measurements to accomplish this task, but SICs use generalized measurements that involve coupling the system to an ancilla and making measurements on the ancilla.\\

The purpose of this paper is to show how the Majorana description of a spin-1 system can be used to deduce its MUBs and SICs. The argument is based entirely on an expression for the overlap of two spin-1 states in terms of their Majorana vectors, given in Eq.(\ref{olap1}) below. The treatment thus has an elementary character that may make it appealing to readers with only a limited background of quantum mechanics. Perhaps the best way of setting the stage for the derivations to be presented below is to show how a similar approach can be used to deduce the MUBs and SICs of a spin-half system.\\

An arbitrary (pure) state of a spin-half particle can be represented by a point on the unit sphere and written as $|\vec{a}\rangle$, where $\vec{a}$ is the unit vector corresponding to the point on the sphere. The ``overlap'' of the states $|\vec{a}\rangle$ and $|\vec{b}\rangle$, defined as $|\langle \vec{b}|\vec{a}\rangle|^{2}$, is given by

\begin{equation}
|\langle \vec{b}|\vec{a}\rangle|^{2} = \frac{1}{2}(1+\vec{a}\cdot\vec{b}).
\label{olap1}
\end{equation}
\noindent
Two states are said to orthogonal, unbiased or equiangular{\footnote{The term equiangular seems apt since the states of a SIC are often spoken of as a set of equiangular lines.}} if their overlap is 0, $1/2$ or $1/3$, respectively. Thus two states are orthogonal if their vectors are oppositely directed, unbiased if their vectors are perpendicular and equiangular if their vectors make an angle of $\cos^{-1}(-\frac{1}{3})$ with each other. The problem we take up now is that of using these definitions and purely geometrical arguments to deduce the MUBs and SICs of a spin-half system.\\

Let us begin with the MUBs. The states of any basis are represented by diametrically opposite points on the unit sphere, and a pair of mutually unbiased bases are represented by the points at the ends of two perpendicular diameters. However a sphere can have no more than three mutually perpendicular diameters, and so it follows that the maximum number of mutually unbiased bases is three. The points representing the states of the bases lie at the vertices of a regular octahedron, and rotating the octahedron rigidly about its center yields an infinite family of MUBs all of whose members are unitarily equivalent to each other.\\

Next consider the SICs. The equiangularity condition implies that the unit vectors corresponding to two equiangular states make an angle of $\cos^{-1}(-\frac{1}{3})$ with each other. However this is just the angle subtended by an edge of a regular tetrahedron at its center, and it follows that a SIC can consist of at most four states whose representative points lie at the vertices of a regular tetrahedron. Rotating this tetrahedron about its center yields an infinite family of SICs all of whose members are unitarily equivalent to each other.\\

The problem addressed in this paper is that of generalizing the above arguments to a spin-1 system. The task is now harder because a state of a spin-1 system is represented by a pair of unit vectors and the formula for the overlap is more complicated. We present the needed formula in Sec.2 and use it to derive the conditions that the vectors of a pair of spin-1 states must satisfy if they are to be orthogonal, unbiased or equiangular. In Sec.3 we show how the orthogonality and unbiasedness conditions imply the existence of a set of four mutually unbiased bases, which is unique up to unitaries. In Sec.4 we show how the equiangularity condition can be used as a starting point for the derivation of a potentially large number of SICs; however technical difficulties make it possible to push through the construction in just two cases, both of which yield SICs that are well known form earlier work. Finally, Sec.5 tries to convey the new light that this approach sheds on the MUBs and the SICs, and particularly its limitations in uncovering all the known SICs.

\section{\label{sec:2} Overlap formula for spin-1 states}

In the Majorana approach \cite{majorana}-\cite{hannay}, a pure state of a spin-1 system is represented by a pair of unit vectors that we will refer to as its Majorana vectors, or M-vectors for short. We will write a state as $|\vec{a_{1}},\vec{a_{2}}\rangle$, where $\vec{a_{1}}$ and $\vec{a_{2}}$ are its (unordered) M-vectors. We distinguish between three types of states (see Fig.1): those whose M-vectors are parallel, which we will refer to as coherent states or C-states; those whose M-vectors are antiparallel, which we will refer to as anticoherent states or A-states; and those whose M-vectors make an arbitrary angle with each other, which we will refer to as devious states or D-states. Of course C- and A-states are just special cases of D-states, and all these states can be turned into each other by SU(3) transformations, but we will still find it to be of value to maintain this distinction in the arguments to be presented below.\\

\begin{figure}[htp]
\begin{center}
\includegraphics[width=.30\textwidth]{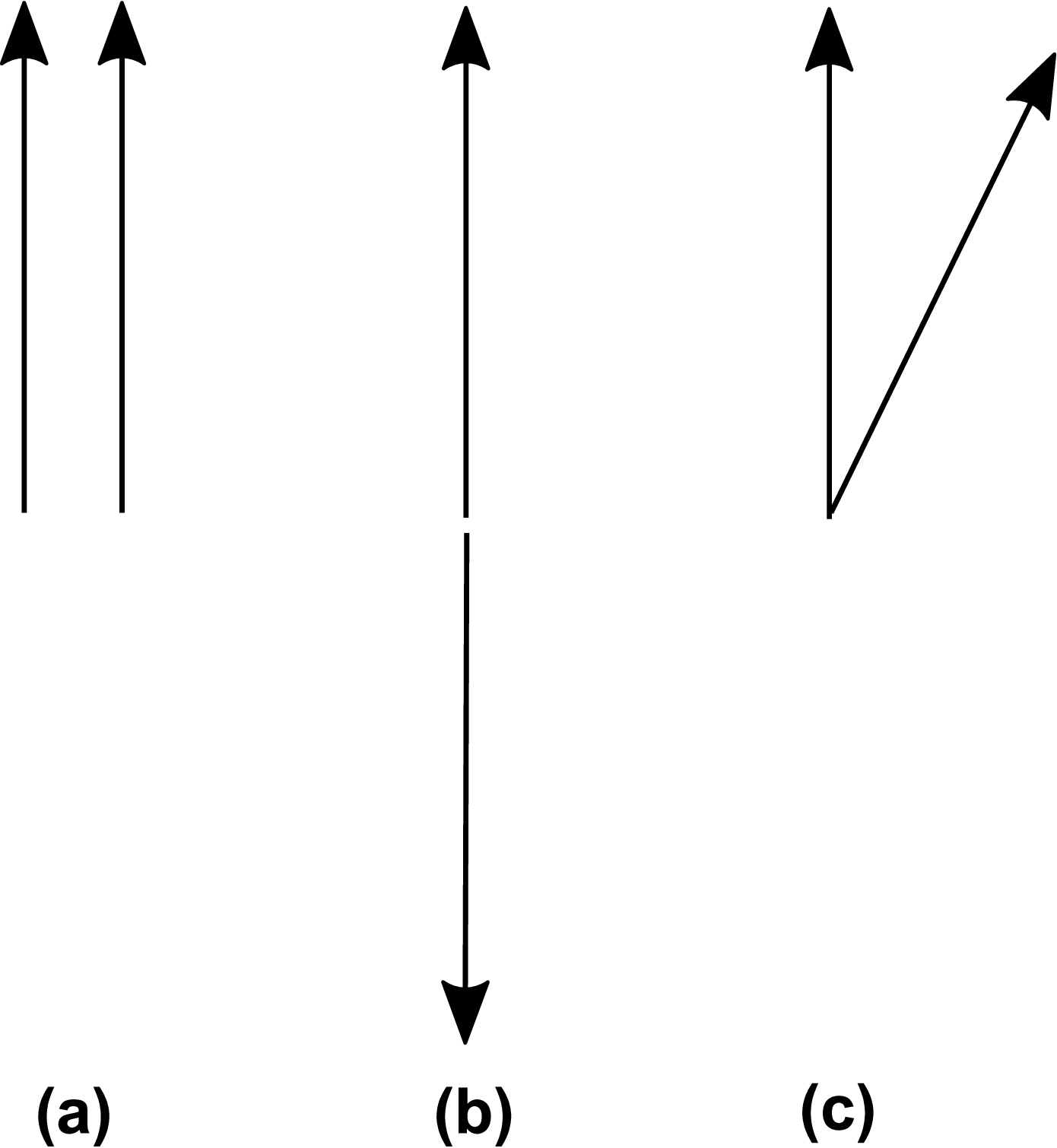}
\end{center}
\caption{The three different types of spin-1 states: (a) Coherent or C-state, with parallel M-vectors, (b) Anticoherent or A-state, with antiparallel M-vectors, and (c) Devious or D-state, with M-vectors making an arbitrary angle with each other.}
\label{Fig.1}
\end{figure}
 The main tool we will need in this paper is an expression for the overlap of two spin-1 states in terms of their M-vectors. It is{\footnote{While such a formula will not come as a surprise to many, published references to it are hard to find. To our knowledge, it was first published in \cite{gould}.}}

\begin{equation}
|\langle \vec{b_{1}},\vec{b_{2}} | \vec{a_{1}},\vec{a_{2}} \rangle|^{2} = \frac{2F-(1-\vec{a_{1}}\cdot\vec{a_{2}})(1-\vec{b_{1}}\cdot\vec{b_{2}})} {(3+\vec{a_{1}}\cdot\vec{a_{2}})(3+\vec{b_{1}}\cdot\vec{b_{2}})} \hspace{2mm} ,
\label{olap2}
\end{equation}
\\
\begin{equation}
\text{where} \hspace{6mm} F=(1+\vec{a_{1}}\cdot\vec{b_{1}})(1+\vec{a_{2}}\cdot\vec{b_{2}})+(1+\vec{a_{1}}\cdot\vec{b_{2}})(1+\vec{a_{2}}\cdot\vec{b_{1}}).\\
\label{olap2ext}
\end{equation}
\\ \noindent
A derivation of this formula is given in Appendix 1. Two spin-1 states are said to be identical, orthogonal, unbiased or equiangular if their overlap is 1,0, $\frac{1}{3}$ or $\frac{1}{4}$, respectively. Using (\ref{olap2}), these conditions can be expressed in terms of the M-vectors of the states as\\
\begin{equation}
\text{Identity:} \hspace{23mm}  F-(1+\vec{a_{1}}\cdot\vec{a_{2}})(1+\vec{b_{1}}\cdot\vec{b_{2}})-4=0
\label{iden}
\end{equation}
\noindent
\begin{equation}
\text{Orthogonality:} \hspace{32mm}  2F-(1-\vec{a_{1}}\cdot\vec{a_{2}})(1-\vec{b_{1}}\cdot\vec{b_{2}})=0
\label{orth}
\end{equation}
\noindent
\begin{equation}
\text{Unbiasedness:} \hspace{35mm}  3F-2(\vec{a_{1}}\cdot\vec{a_{2}})(\vec{b_{1}}\cdot\vec{b_{2}})-6=0
\label{unb}
\end{equation}
\noindent
\begin{equation}
\text{Equiangularity:} \hspace{10mm}  8F-5(\vec{a_{1}}\cdot\vec{a_{2}})(\vec{b_{1}}\cdot\vec{b_{2}})+\vec{a_{1}}\cdot\vec{a_{2}}+\vec{b_{1}}\cdot\vec{b_{2}}-13=0
\label{equi}
\end{equation}
\\ \noindent
Two states satisfy the condition for being identical, Eq.(\ref{iden}), if and only if $\vec{a_{1}}$ and $\vec{a_{2}}$ are equal to $\vec{b_{1}}$ and $\vec{b_{2}}$ (in either order). The task now is to use (\ref{iden})-(\ref{equi}) to deduce the MUBs and SICs of a spin-1 system.\\

We will often find it convenient, in the arguments below, to refer to an M-vector $\vec{a}$ by its spherical coordinates{\footnote{To be clear, the Cartesian coordinates of $\vec{a}$ are $(\sin\theta\cos\phi,\sin\theta\sin\phi,\cos\theta)$.} $\theta,\phi$. Correspondingly the state $|\vec{a_{1}},\vec{a_{2}}\rangle$ will be written as $(\theta_{1},\phi_{1}|\theta_{2},\phi_{2})$, with the round brackets replacing the ket to indicate that this is a state expressed in angular notation.

\section{\label{sec:3} MUBs of a spin-1 system}

We present the argument leading to the MUBs{\footnote{We will henceforth use the singular form MUB to refer to a single set of mutually unbiased bases and the plural MUBs to refer to several such sets.}} in a number of steps.\\

\subsection{\label{subsec:3.1} \bf{Initial basis}}

We take as the first basis of a MUB the CAC basis consisting of the C-state\footnote{The azimuthal angles of all the states in this basis are undefined, but we have put them equal to 0 for simplicity.} $|\vec{z},\vec{z}\rangle = (0,0|0,0)$, the A-state $|\vec{z},-\vec{z}\rangle = (0,0|\pi,0)$ and the C-state $|-\vec{z},-\vec{z}\rangle = (\pi,0|\pi,0)$, where $\vec{z}$ is a unit vector along the positive z-axis (see Fig.2). This is the standard angular momentum basis, with the states having a spin component $+1$, $0$ or $-1$ along the z-axis. There is no loss of generality in this choice of initial basis, because any other basis can always be brought into this form by a suitable SU(3) transformation.

\begin{figure}[htp]
\begin{center}
\includegraphics[width=.30\textwidth]{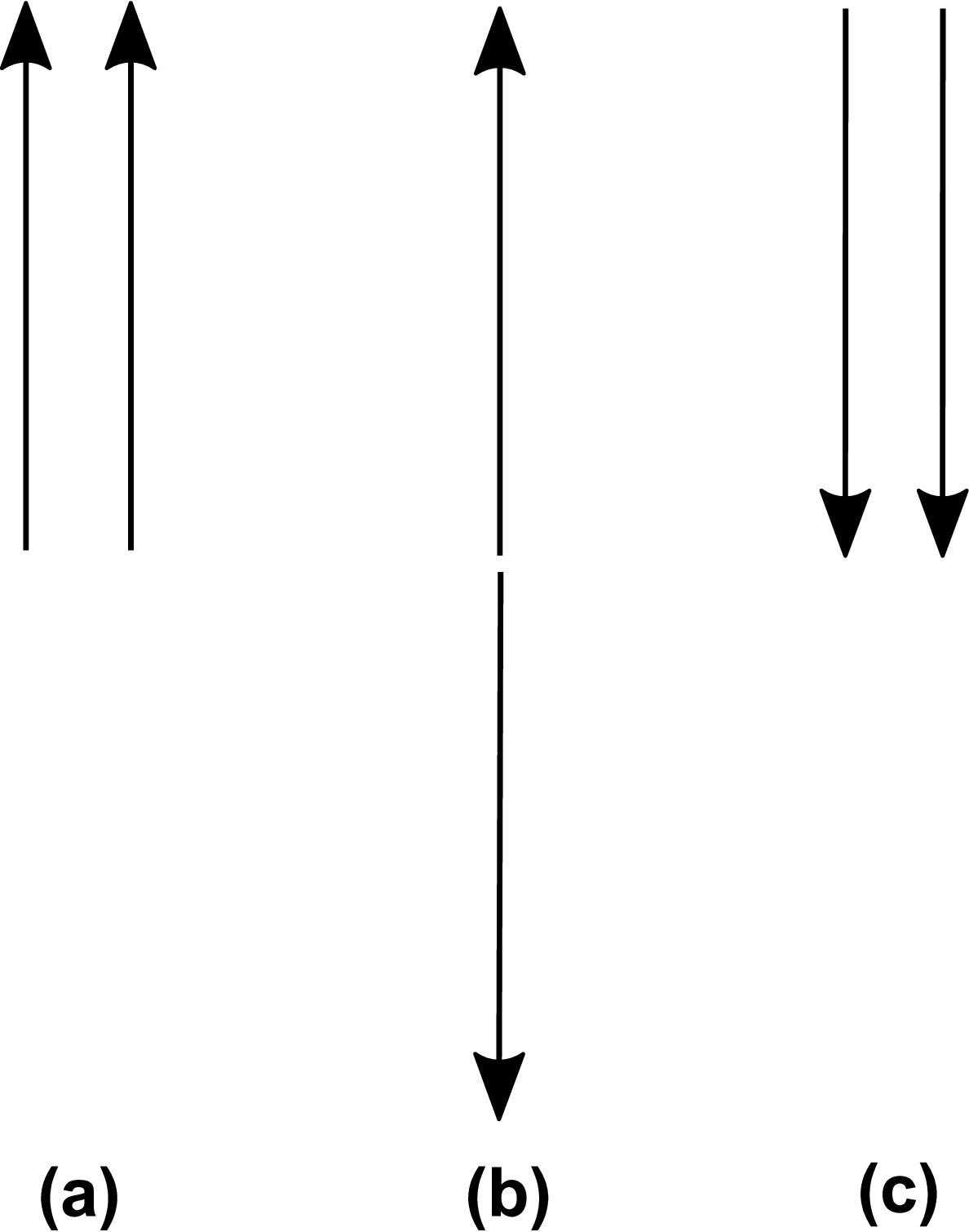}
\end{center}
\caption{The CAC basis, consisting of (a) a C-state with both M-vectors pointing up, (b) an A-state with one vector pointing up and the other down, and (c) a second C-state with both vectors pointing down. This is the standard angular momentum basis, with the states having a spin-component of $+1,0$ or $-1$ along the vertical or z-direction. This basis is also the first of the four bases of the MUBs constructed in this paper.}
\label{Fig.2}
\end{figure}

\subsection{\label{subsec:3.2} \bf{Double-cone states}}

All the other states of the MUB must be unbiased to each of the states of the CAC basis, and we will see that this implies a strong constraint on the form they can have. If $|\vec{a_{1}},\vec{a_{2}}\rangle$ denotes one of these other states, then the fact that it must be unbiased to each of the states of the CAC basis leads, via (\ref{unb}), to the three equations
\begin{align}
\label{eqn:unb3}
\begin{split}
 3(a_{1z}+a_{2z}+a_{1z}a_{2z})=\vec{a_{1}}\cdot \vec{a_{2}} ,
\\
 3(-a_{1z}-a_{2z}+a_{1z}a_{2z})=\vec{a_{1}}\cdot \vec{a_{2}} ,
 \\
\text{and} \hspace{35mm}  3a_{1z}a_{2z}=\vec{a_{1}}\cdot \vec{a_{2}} ,
\end{split}
\end{align}

\noindent
where $a_{iz}=\vec{z}\cdot\vec{a_{i}}$ $(i=1,2)$. These equations imply that $a_{2z}=-a_{1z}$ and $\vec{a_{1}}\cdot \vec{a_{2}}=-3a_{1z}^{2}$; in other words, they imply that the vectors $\vec{a_{1}}$ and $\vec{a_{2}}$ lie on the upper and lower halves of a double-cone of semi-vertex angle $\theta$ (say) and make an angle of $\cos^{-1}(-3\cos^{2}\theta)$ with each other, where $\cos^{-1}(\frac{1}{\sqrt{3}}) \leq \theta \leq \frac{\pi}{2}$ (see Fig.\ref{Fig.3}). We will refer to these states as ``double-cone'' states. There are actually two different types of double-cone states on any double-cone and they can specified in terms of their angular coordinates as
\begin{align}
\label{DCdef}
\begin{split}
 D^{R}(\theta,\phi) \equiv (\theta,\phi|\pi-\theta,\phi+\pi-\phi_{\theta})
\\
\text{and} \hspace{6mm} D^{L}(\theta,\phi) \equiv (\theta,\phi|\pi-\theta,\phi+\pi+\phi_{\theta}) \hspace{2mm} ,
\end{split}
\end{align}

\noindent
where $\phi_{\theta} = \cos^{-1}(2\cot^{2}\theta)$ is an acute angle. We will refer to $\phi_{\theta}$ as the ``twist angle'' because it is the angle by which the lower vector is rotated on its double-cone from its straight line position with the upper vector. This rotation is in the clockwise sense for the state $D^{R}$ and the counter-clockwise sense for $D^{L}$; thus these states have different chiralities, as noted in their superscripts $R$ and $L$.  The states $D^{R}(\theta,\phi)$ and the $D^{L}(\theta,\phi)$ both form a continuous family for $0\leq \phi < 2\pi$, but no member of one coincides with any member of the other.\\

\noindent
The double-cone states are particular examples of D-states, and they are in fact the only types of D-states we ever have to consider in connection with MUBs. The double-cone states have some important properties that we state in the form of a number of propositions:\\

\noindent
{\bf Proposition 1}. Let $\Theta$ be the operation of inverting the M-vectors of a state in the origin\footnote{The physical significance of this operation is that it corresponds to time-reversal.}. Then $\Theta D^{R}(\theta,\phi) = D^{L}(\theta,\phi-\phi_{\theta})$. \\

\noindent
Proof: This is easily seen from (\ref{DCdef}) on noting that $\theta\rightarrow\pi-\theta$ and $\phi \rightarrow \phi+\pi$ under inversion and that the order of the arguments of a double-cone state is irrelevant. \hspace{2mm} $\Box$ \\

\noindent
In words, this relation says that inverting the vectors of $D^{R}(\theta,\phi)$ in the origin produces the same state that would be obtained by rotating the vectors of $D^{L}(\theta,\phi)$ clockwise about the z-axis by the angle $\phi_{\theta}$.\\

\noindent
{\bf Proposition 2}. Double-cone states are chiral for $\cos^{-1}(1/\sqrt{3}) < \theta < \pi/2$ but lose their chirality at the two limits of this range.\\

\noindent
Proof: The chirality within this range was already pointed out earlier, so we just need to look at the two limits. At the lower limit, $\theta = \cos^{-1}(1/\sqrt{3})= 54.74^{\circ}$, the twist angle goes to zero and the state degenerates into an A-state, which is not chiral. And at the upper limit, $\theta = \pi/2$ or $3\pi/2$, and the double-cone flattens out into a plane with the M-vectors becoming orthogonal; the chirality is again lost because the inverse of a state in the origin can be obtained by rotating the vectors of the original state by a half-turn about the z-axis. We will dispense with the superscripts $R$ and $L$ for these two limiting cases of the double-cone states. \hspace{2mm} $\Box$ \\

\noindent
{\bf Proposition 3}. The overlap of the states $D^{R}(\theta,0)$ and $D^{R}(\theta,\phi)$ vanishes only for $\phi = 2\pi/3$ or $4\pi/3$. The same is true of the $L$ states.\\

\noindent
Proof: On using (\ref{orth}), the orthogonality condition for a pair of $R$ or $L$ states takes the form

\begin{equation}\label{RLorth}
(2\cos\phi+1)^{2}\sin^{4}\theta=0 \hspace{2mm},
\end{equation}

\noindent
from which the result follows immediately.\hspace{2mm} $\Box$\\

\noindent
\textbf{Remark.} Note that the three states $D^{R}(\theta,0)$, $D^{R}(\theta,2\pi/3)$ and $D^{R}(\theta,4\pi/3)$ form a basis, as do the corresponding $L$ states. The double-cone states at the two boundaries of the range, which are not chiral, also have this property.\\

\noindent
{\bf Proposition 4}. The only values of $\phi$ for which the states $D^{R}(\theta,0)$ and $D^{R}(\theta,\phi)$ become unbiased are $\pm \cos^{-1}(\frac{\sqrt{3}-1}{2}) = \pm68.53^{\circ}$. The same is true of the $L$ states.\\

\noindent
Proof: On using (\ref{unb}), the unbiasedness condition for either the $R$ or $L$ states takes the form

\begin{equation}\label{RLunb}
(2\cos^{2}\phi+2\cos\phi-1)\sin^{4}\theta=0 \hspace{2mm} ,
\end{equation}

\noindent
from which the result follows immediately. \hspace{2mm} $\Box$

\begin{figure}[htp]
\begin{center}
\includegraphics[width=.40\textwidth]{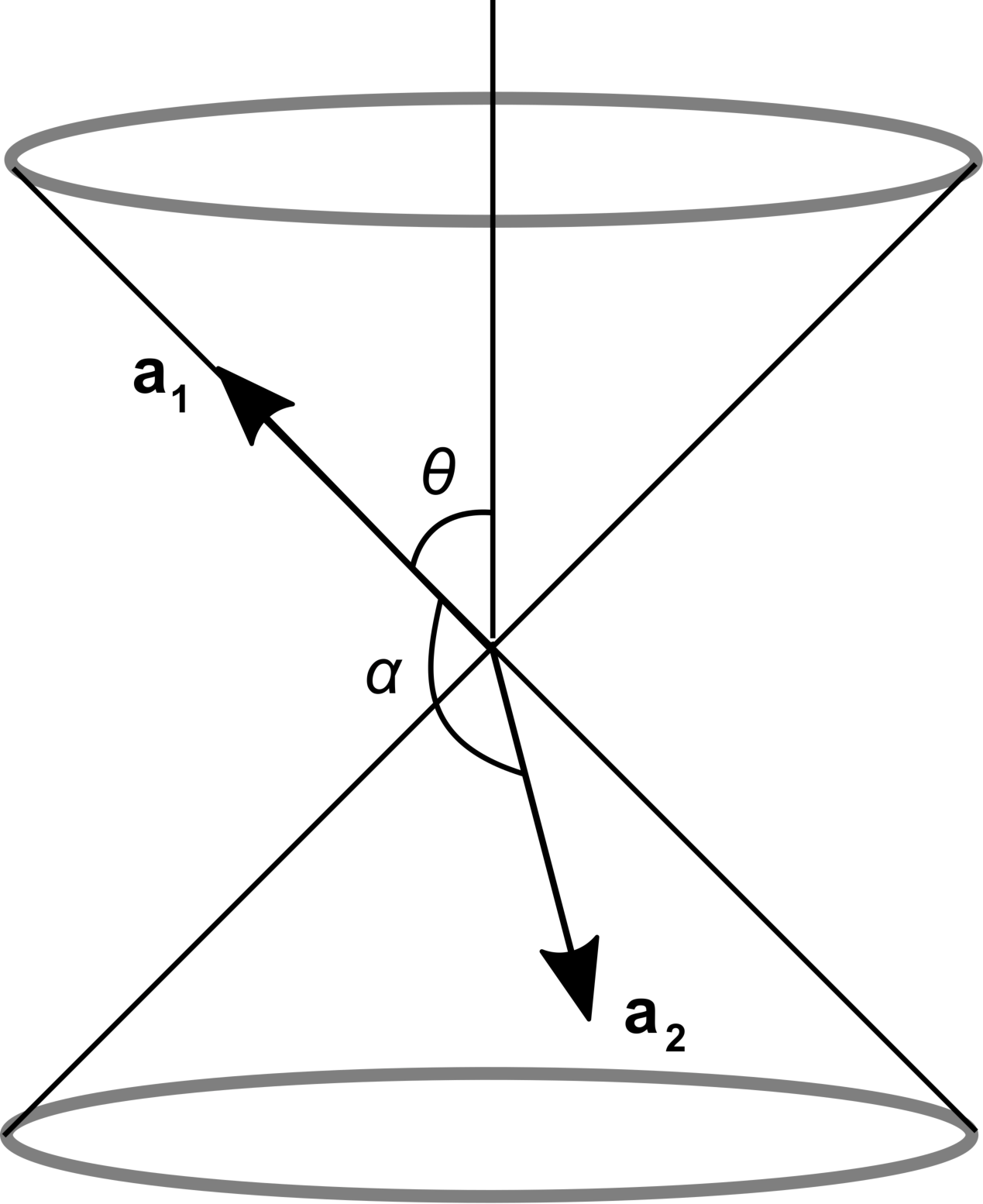}
\end{center}
\caption{A double-cone state, with one M-vector lying on the upper half of a double-cone of semi-vertex angle $\theta$ and the other lying on the lower half. The angle between the two M-vectors, labeled $\alpha$, is $\cos^{-1}(-3\cos^{2}\theta)$. The lower vector is rotated on its cone by the twist angle $\phi_{\theta}=\cos^{1}(-2\cot^{2}\theta)$ from its straight line position with the upper vector, with the rotation being clockwise for a $R$ state or counter-clockwise for a $L$ states. Rotating this state by $120^{\circ}$ or $240^{\circ}$ about the axis of the double-cone yields two other states that form a basis along with the original state.}
\label{Fig.3}
\end{figure}

\subsection{\label{subsec:3.3} \bf{States on different double-cones}}

So far we have looked at only states lying on the same double-cone. However we must now consider the relationship between states lying on different double-cones, as that will prove crucial in the construction of a MUB. Let us consider the state $D^{R,L}(\theta_{1},0)$ on the cone $\theta_{1}$ and the state $D^{R,L}(\theta_{2},\phi)$ on the cone $\theta_{2}$, with the latter being rotated azimuthally by the angle $\phi$ relative to the first (the superscripts $R,L$ mean that either state can be a $R$ or $L$ state; all the cases can be treated simultaneously by allowing the twist angles $\phi_{1}$ and $\phi_{2}$ of the two states\footnote{The twist angle associated with the state $D^{R,L}(\theta_{1},0)$ should be written $\phi_{\theta_{1}}$, but this is rather cumbersome so we will abbreviate it to $\phi_{1}$ and use similar contractions for the other twist angles that occur.} to assume positive or negative values for a $R$ or $L$ state, respectively). The basic question we want to answer is this: when can the states $D^{R,L}(\theta_{1},0)$ and $D^{R,L}(\theta_{2},\phi)$ be orthogonal or unbiased to each other? The propositions of the previous subsection answered this question for $\theta_{1}=\theta_{2}$, but we would now like the answers when $\theta_{1} \neq  \theta_{2}$. The answers can be found by applying the criteria (\ref{orth}) and (\ref{unb}) to these states. After some algebra, the criterion for orthogonality can be written as $F(\phi)+c_{3} = 0$ and that for unbiasedness as $F(\phi)+c_{4} = 0$ where

\begin{equation}
\label{orthunb}
 F(\phi) = \sin\phi(c_{1}+c_{2}\cos{\phi})+ \cos\phi(d_{1}+d_{2}\cos{\phi})
\end{equation}

\noindent
and the quantities $c_{1},c_{2},d_{1},d_{2},c_{3}$ and $c_{4}$ are defined as
\begin{multline}
c_{1}=\sin\theta_{1}\sin\theta_{2}\Big[\cos\theta_{1}\cos\theta_{2}\Big(\sin\phi_{2}-\sin\phi_{1}+\sin(\phi_{2}-\phi_{1})\Big)
\\ + \Big(\sin\phi_{1}-\sin\phi_{2}+\sin(\phi_{2}-\phi_{1})\Big)\Big]
\label{c1}
\end{multline}
\begin{multline}
d_{1}=\sin\theta_{1}\sin\theta_{2}\Big[\cos\theta_{1}\cos\theta_{2}\Big((1+\cos\phi_{1})(1+\cos\phi_{2})+\sin\phi_{1}\sin\phi_{2}\Big)
\\ + \Big((1-\cos\phi_{1})(1-\cos\phi_{2})+\sin\phi_{1}\sin\phi_{2}\Big)\Big]
\label{d1}
\end{multline}
\begin{equation}
c_{2}=2\sin^{2}\theta_{1}\sin^{2}\theta_{2}\sin\big(\phi_{2}-\phi_{1}\big)
\label{c2}
\end{equation}
\begin{equation}
d_{2}=2\sin^{2}\theta_{1}\sin^{2}\theta_{2}\cos\big(\phi_{2}-\phi_{1}\big)
\label{d2}
\end{equation}
\begin{equation}
c_{4}=-4\cos^{2}\theta_{1}\cos^{2}\theta_{2} - \sin^{2}\theta_{1}\sin^{2}\theta_{2}\sin\phi_{1}\sin\phi_{2}
\label{c4}
\end{equation}
\begin{equation}
\text{and} \hspace{5mm} c_{3} =c_{4} + \frac{3}{2}\sin^{2}\theta_{1}\sin^{2}\theta_{2} \hspace{2mm}.
\label{c3}
\end{equation}

\noindent
The function $F(\phi)$ can be cast into a simpler form by the variable change $\phi^{\prime}=\phi-\phi_{0}$, with $\phi_{0}$ being a suitable constant. Then it takes the form

\begin{equation}
F(\phi^{\prime}) = \sin\phi^{\prime}(c^{\prime}_{1}+c^{\prime}_{2}\cos{\phi^{\prime}})+ \cos\phi^{\prime}(d^{\prime}_{1}+d^{\prime}_{2}\cos{\phi^{\prime}})+c^{\prime}_{3}
\label{Fprime}
\end{equation}
\noindent
\begin{equation}\label{c12prime}
\text{where} \hspace{5mm} c^{\prime}_{1}=c_{1}\cos\phi_{0}-d_{1}\sin\phi_{0}  \hspace{2mm} ,  \hspace{2mm} c^{\prime}_{2}=c_{2}\cos2\phi_{0}-d_{2}\sin2\phi_{0}
\end{equation}
\begin{equation}\label{d12prime}
\hspace{5mm} d^{\prime}_{1}=d_{1}\cos\phi_{0}+c_{1}\sin\phi_{0}  \hspace{2mm} ,  \hspace{2mm} d^{\prime}_{2}=d_{2}\cos2\phi_{0}+c_{2}\sin2\phi_{0}
\end{equation}
\begin{equation}\label{c3prime}
\text{and} \hspace{10mm} c^{\prime}_{3}=\frac{1}{2}d_{2}\Big(1-\frac{1}{\cos2\phi_{0}}\Big)
\end{equation}

\noindent
If one chooses $\phi_{0}$ so that $c^{\prime}_{1}=c^{\prime}_{2}=0$, then (\ref{Fprime}) reduces to

\begin{equation}\label{Fprime2}
F(\phi^{\prime}) = \cos\phi^{\prime}(d^{\prime}_{1}+d^{\prime}_{2}\cos{\phi^{\prime}})+c^{\prime}_{3} \hspace{2mm} .
\end{equation}

\noindent
However for this to happen both the conditions

\begin{equation}\label{ph0}
\hspace{5mm} \tan\phi_{0} = \frac{c_{1}}{d_{1}}  \hspace{5mm} \text{and} \hspace{5mm} \tan2\phi_{0} = \frac{c_{2}}{d_{2}}
\end{equation}

\noindent
must be satisfied. These conditions can be satisfied simultaneously only if $d_{1}(c_{2}d_{1}-c_{1}d_{2})=c_{1}(c_{1}c_{2}+d_{1}d_{2})$, but it is easily verified that this is the case and one then sees, with the aid of (\ref{c2}) and (\ref{d2}), that

\begin{equation}\label{ph02}
\phi_{0} = \frac{1}{2}\big(\phi_{2}-\phi_{1}) \hspace{2mm}.
\end{equation}

\noindent
With $\phi_{0}$ thus fixed, the coefficients in (\ref{Fprime2}) assume the values
\begin{multline}
\label{d1p}
d^{\prime}_{1} = \frac{d_{1}}{\cos\phi_{0}} = \pm\sqrt{c_{1}^{2}+d_{1}^{2}} = \pm 2\sin\theta_{1}\sin\theta_{2}\Big[\cos^{2}\theta_{1}\cos^{2}\theta_{2}\big(1+\cos\phi_{1})(1+\cos\phi_{2}\big)
\\ + \big(1-\cos\phi_{1})(1-\cos\phi_{2}\big) +2\cos\theta_{1}\cos\theta_{2}\sin\phi_{1}\sin\phi_{2}\Big]
\end{multline}
\begin{equation}\label{d2p}
d^{\prime}_{2} = \frac{d_{2}}{\cos2\phi_{0}} =2\sin^{2}\theta_{1}\sin^{2}\theta_{2}
\end{equation}
\begin{equation}\label{c3p}
\hspace{2mm} \text{and} \hspace{2mm} c^{\prime}_{3} = \sin^{2}\theta_{1}\sin^{2}\theta_{2}(\cos2\phi_{0}-1) \hspace{2mm}.
\end{equation}

\noindent
Eqs.(\ref{Fprime2}),(\ref{d1p}),(\ref{d2p}) and (\ref{c3p}) are the form of the function $F$ that we will employ in the subsequent analysis. This form is particularly convenient because it reveals $F$ to be an even function of $\phi^{\prime}$ with two unequal maxima and two equal minima (see Fig.4). The two propositions to be proved now can be appreciated better in the context of this figure. \\

\begin{figure}[htp]
\begin{center}
\includegraphics[width=.40\textwidth]{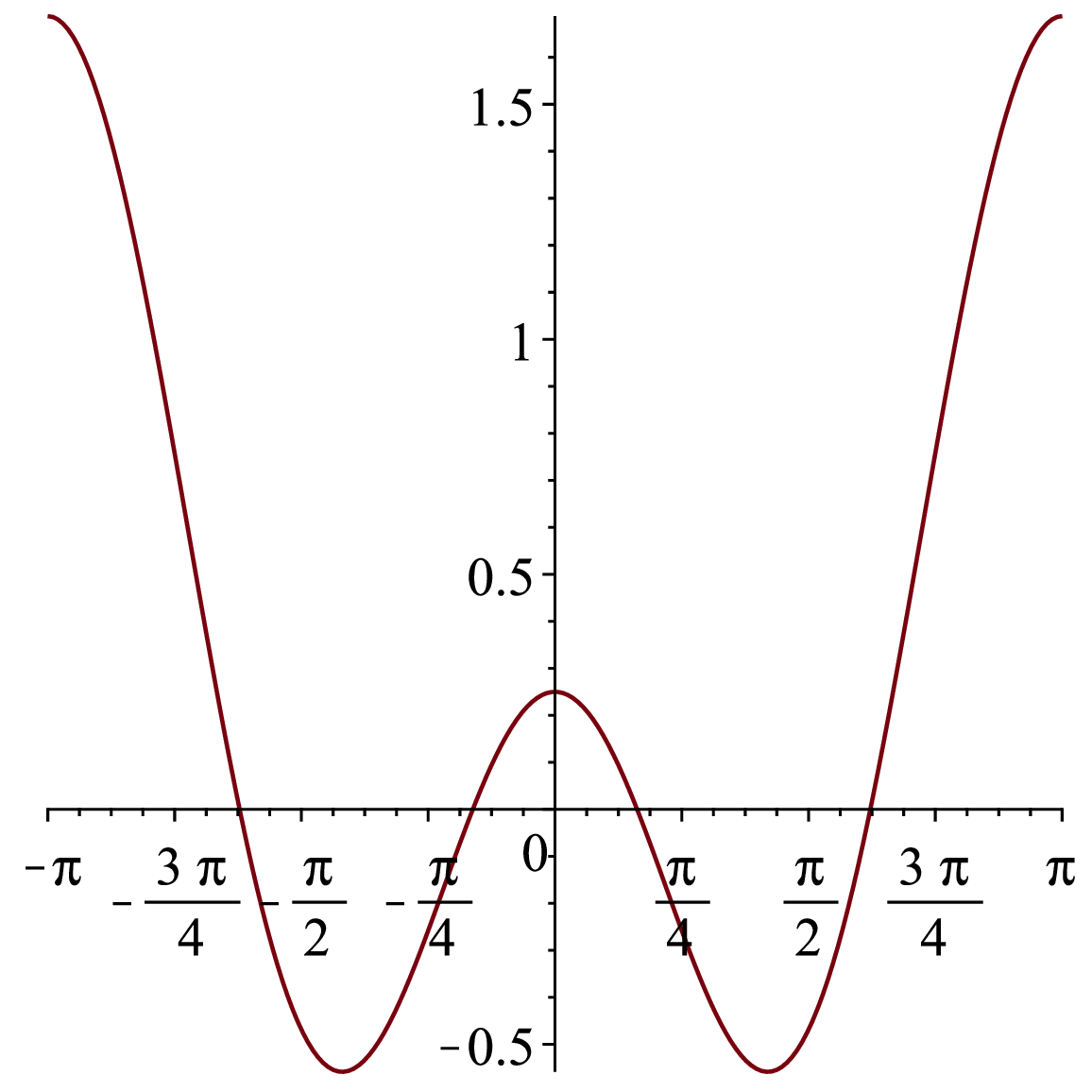}
\end{center}
\caption{Plot of the function $F(\phi^{\prime})$ of Eq.(\ref{Fprime2}) over a complete period. It is an even function in the shape of a W, with two unequal maxima and two equal minima. The minima occur symmetrically about the maximum at the origin. The conditions that determine the orthogonality and unbiasedness of states lying on the double-cones $\theta_{1}$ and $\theta_{2}$ depend on how this function is shifted up or down along the vertical direction by the added constant $c_{3}$ or $c_{4}$.}
\label{FigF}
\end{figure}

\noindent
{\bf Proposition 5}. Two states on different double-cones can never be orthogonal.\\

\noindent
Proof: We must show that $F(\phi^{\prime}) + c_{3} > 0$ for all $\phi^{\prime}$ if $\theta_{1} \neq \theta_{2}$. On putting the derivative of $F$ with respect to $\phi^{\prime}$ equal to zero we find that the maxima of $F$ occur at $\phi^{\prime} = 0$ and $\pi$ and the minima at $\phi^{\prime} = \cos^{-1}\Big(-d^{\prime}_{1}/2d^{\prime}_{2} \Big)$. The value of $F(\phi^{\prime}) + c_{3}$ at the minimum is $-\frac{{d^{\prime}_{1}}^{2}}{4d^{\prime}_{2}} +  c^{\prime}_{3} +c_{3}$, and using (\ref{c4}),(\ref{c3}),(\ref{d1p}),(\ref{d2p}) and (\ref{c3p}) allows the value of the minimum to be written as

\begin{multline}\label{orthcon}
\frac{1}{2}\Big[-\cos^{2}\theta_{1}\cos^{2}\theta_{2}\big(\cos\phi_{1}+\cos\phi_{2}+\cos\phi_{1}\cos\phi_{2}\big) \\ +\big(\cos\phi_{1}+\cos\phi_{2}-\cos\phi_{1}\cos\phi_{2}\big) -2\cos\theta_{1}\cos\theta_{2}\sin\phi_{1}\sin\phi_{2}\Big]
\end{multline}

\noindent
We must now show that this expression is positive definite for $\theta_{1} \neq \theta_{2}$. However this follows from the fact that (\ref{orthcon}) is symmetrical in the parameters of the two states and so achieves its extreme (here minimum) value when $\theta_{1} = \theta_{2}$ and $\phi_{1} = \phi_{2}$; but this minimum value is zero, and so it follows that (\ref{orthcon}) is positive definite for $\theta_{1} \neq \theta_{2}$, which proves the claim. \hspace{2mm} $\Box$\\

\noindent
\textbf{Remark 1.} The geometrical meaning of this proposition can be understood by looking at Fig.4: essentially, adding the constant $c_{3}$ to the function $F(\phi^{\prime})$ raises it vertically so that both its (equal) minima lie above the $x$-axis.\\

\noindent
\textbf{Remark 2 (a corollary).} When $\theta_{1} = \theta_{2}$, the two minima of $F$ just touch the $x$-axis at
$\phi^{\prime} = \pm 2\pi/3$ and, in addition, $\phi_{0} = (\phi_{2}-\phi_{1})/2=0$; together these results imply that the states orthogonal to a given state lie on the on the same double-cone as it and are rotated from it by angles of $\pm 2\pi/3$ about the cone axis, which is just the result proved in Proposition 3. However we are now in a position to make the stronger statement that any double-cone state is a member of a \emph{unique} basis consisting of it and the two other states obtained from it by rotations of $\pm 2\pi/3$ about the cone axis. We will use the notation $\big[D^{R,L}(\theta,\phi)\big]$ for such a basis, identifying it by its first member, with the understanding that the other two members are $D^{R,L}(\theta,\phi+2\pi/3)$ and $D^{R,L}(\theta,\phi+4\pi/3)$.\\

\noindent
\textbf{Remark 3.} A much quicker proof of this proposition can be given if we represent the states as rays in Hilbert space. Let $1$ be a state on the double-cone $\theta_{1}$ and $2$ a state on a different double-cone $\theta_{2}$ that is assumed to be orthogonal to 1. We can represent $1$ and $2$ by the rays $(1,0,0)$ and $(0,1,0)$. Now rotate both $1$ and $2$ on their double-cones by the angle $2\pi/3$ so that they go into the states $1^{\prime}$ and $2^{\prime}$. But, from Proposition 3, $1^{\prime}$ is orthogonal to $1$ and $2^{\prime}$ to $2$, so these states can be represented as $1^{\prime}=(0,\alpha_{1},\alpha_{2})$ and $2^{\prime}=(\beta_{1},0,\beta_{2})$. However the M-vectors of $2^{\prime}$ and $1^{\prime}$ bear the same relationship to each other as those of $2$ and $1$, and so $2^{\prime}$ and $1^{\prime}$ must be orthogonal. But the inner product of $1^{\prime}$ and $2^{\prime}$ is ${\beta_{2}}^{*}\alpha_{2}$, which can be zero only if $\alpha_{2}=0$ or $\beta_{2}=0$; however the former choice makes $1^{\prime}=2$ and the latter makes $2^{\prime}=1$, neither of which is permissible, so it follows that $1$ and $2$ cannot be orthogonal, as originally assumed.\\

\noindent
Let us pause to take stock of where we are at. We chose the initial basis of the MUB to be a CAC basis and we have now shown, as a corollary of Proposition 5, that all the remaining bases must be double-cone bases of the form $\big[D^{R,L}(\theta,\phi)\big]$. It remains to see how many bases of this kind we can pick that will form a MUB together with the original CAC basis. The answer to this question is given in the next proposition.\\

\noindent
{\bf Proposition 6}. Let $\big[D^{R}(\theta_{1},0)\big]$ be a double-cone basis, for an arbitrary value of $\theta_{1}$. Then it is always possible to find two other bases, lying on the double-cones $\theta_{2}$ and $\theta_{3}$ that, together with $\big[D^{R}(\theta_{1},0)\big]$ and the CAC basis, form a MUB. To fix $\theta_{2}$ and $\theta_{3}$, let $y=\cos^{2}\theta_{1}$ and consider the quartic equation

\begin{multline}\label{quartic}
(3y^2+6y-1)^2x^4+(36y^4-288y^3-24y^2+20)x^3+(30y^4-24y^3+820y^2-536y+94)x^2
\\ +(-12y^4-536y^2+352y-60)x+(y^2+10y-3)^2=0 \hspace{2mm}.
\end{multline}

\noindent
This equation always has two real roots, $x_2=\cos^{2}\theta_{2}$ and $x_3=\cos^{2}\theta_{3}$, that fix $\theta_{2}$ and $\theta_{3}$. To determine the bases on these double cones, consider the equation

\begin{multline}\label{offset}
\cos\theta_{1}\cos\theta_{j}(1+\cos\phi_{1})(1-\cos\phi_{j})+(1-\cos\phi_{1})(1-\cos\phi_{j})
\\ =\frac{\sigma_{1}\sin\theta_{1}\sin\theta_{2}}{\sqrt{2}}\sqrt{1+\cos\phi_{1}\cos\phi_{j}+\sigma_{2}\sin\phi_{1}\sin\phi_{j}}
 -\sigma_{2}(1+\cos\theta_{1}\cos\theta_{j})\sin\phi_{1}\sin\phi_{j}
\end{multline}

\noindent
where $j=2$ or $3$, $\phi_{1},\phi_{2}$ and $\phi_{3}$ are the (acute) twist angles associated with the double-cones $\theta_{1},\theta_{2}$ and $\theta_{3}$ and $\sigma_{1}$ and $\sigma_{2}$ can each take on the values $+1$ and $-1$. For a given $j$, one can use the value of $\theta_{j}$ found from (\ref{quartic}) to solve (\ref{offset}) for $\phi_{j}$. However one finds that there is a solution for only one combination of values of $\sigma_{1}$ and $\sigma_{2}$ and that it determines a basis on the double-cone $\theta_{j}$ in the manner indicated in the following table:

\begin{table}[ht]
\centering 
\begin{tabular}{|c | c | c | c |} 
\hline 
   &   &    \\
 $\sigma_{1}$ & $\sigma_{2}$ & Basis   \\
  &   &   \\
\hline
 &   &   \\
 $+1$  & $+1$  &$\Big[D^{R}\big(\theta_{j},\phi_{0}+\frac{\pi}{3}\big)\Big]$    \\
  &   &   \\
\hline
   &   &    \\
 $+1$  &$-1$   & $\Big[D^{L}\big(\theta_{j},\phi_{0}+\frac{\pi}{3}\big)\Big]$   \\
   &   &    \\
\hline
   &   &    \\
 $-1$  & $+1$  & $\Big[D^{R}\big(\theta_{j},\phi_{0}\big)\Big]$  \\
   &   &    \\
\hline
   &   &    \\
$-1$  &  $-1$ &  $\Big[D^{L}\big(\theta_{j},\phi_{0}\big)\Big]$  \\
   &   &    \\
\hline
\end{tabular}
\caption{The different solutions of Eq.(\ref{offset}).}
\label{tab0} 
\end{table}
\noindent
The phase offset $\phi_{0}$ in Table \ref{tab0} is given by (\ref{ph02}), which in the present case can be written as $\phi_{0} = \frac{1}{2}\big(\phi_{j}-\phi_{1})$, where $\phi_{j}$ is the twist angle associated with the double-cone $\theta_{j}$.\\

\noindent
Admittedly this result is not simple. However it has the virtue of delivering the entire MUB resulting from the choice of an initial double-cone state (on the cone $\theta_{1}$) in a single, easily implemented, package. Deferring the proof of this proposition to Appendix 2, we concentrate here on trying to convey a feeling for it and then unpack some of the special cases contained in it.\\

\noindent
The quartic equation (\ref{quartic}) that gives rise to the MUB may seem rather complicated at first, but it has an underlying simplicity that allows one to grasp the nature of its solutions. A close inspection shows that it has the form $f(x,y)=0$, where $f(x,y)$ is a symmetrical function of $x$ and $y$. A contour plot of (\ref{quartic}) is shown in Fig.\ref{quarticplot} over the entire region $0\leq x,y \leq \frac{1}{3}$ in which double-cones exist. It is seen to have the shape of a fat boomerang that is symmetrical about the line $y=x$. If we take $y=\cos^{2}\theta_{1}$ and draw a horizontal line corresponding to this value of $y$, it cuts the curve at two points that determine the other two double-cones, $\theta_{2}$ and $\theta_{3}$, of the MUB. However, if $y=0$ or $\frac{1}{3}$, the line is tangent to the curve at a single point that leads to only one additional double-cone. \\

\begin{figure}[htp]
\begin{center}
\includegraphics[width=.40\textwidth]{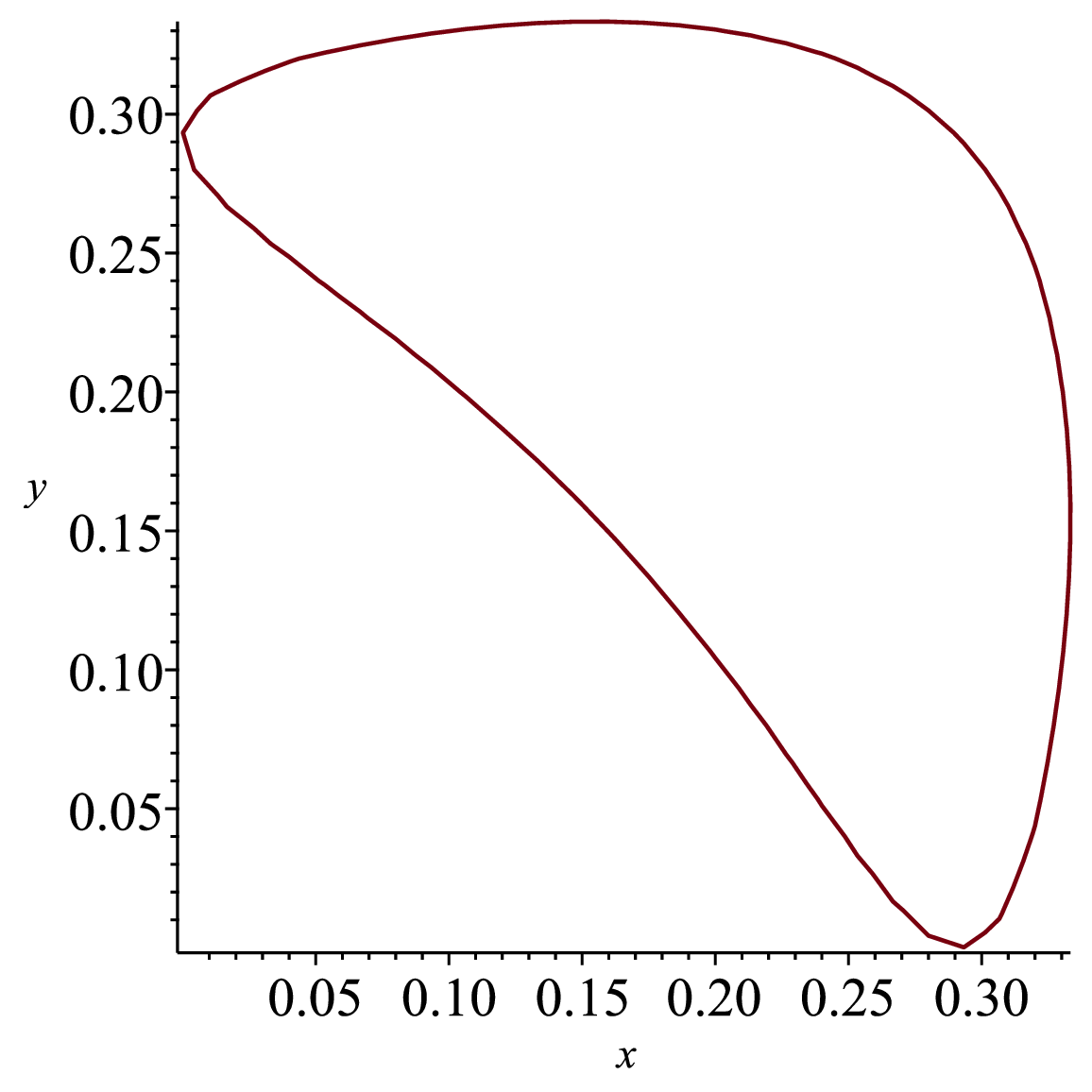}
\end{center}
\caption{Contour plot of the quartic equation (\ref{quartic}) over the entire region in which it is defined. The plot is symmetrical about the line $y=x$. Any line $y$ = constant, which corresponds to a double-cone $\theta_{1}$, generally intersects it in two points, $x_2=\cos^{2}\theta_{2}$ and $x_3=\cos^{2}\theta_{3}$, which determine the other two double-cones of a MUB. However, if $y=0$ or $1/3$, the line is tangent to the boomerang and intersects it in just a single point; in either of these cases the MUB consists of just two double-cones and is not chiral.}
\label{quarticplot}
\end{figure}

\noindent
The MUBs with three double-cones are chiral and have a mix of R and L type bases. Inverting the M-vectors of any such MUB in the origin leads to another MUB of the opposite handedness, and the two together form an enantiomorphic pair. However the special MUBs consisting of just two double-cones are not chiral, as we now demonstrate.\\

\noindent
Consider first the MUB defined by the condition $y=1/3$, for which the quartic reduces to

\begin{equation}\label{caseA}
(3x^2+6x-1)^2=0 \hspace{2mm}.
\end{equation}

\noindent
This equation has a pair of double roots, $\frac{2\sqrt{3}}{3}-1$ and $-\frac{2\sqrt{3}}{3}-1$, of which only the former is positive and acceptable. Thus this MUB has two double-cones of angles $\theta_{a}=\cos^{-1}(\sqrt{\frac{2}{\sqrt{3}}-1})$ and $\theta_{b}=\cos^{-1}(1/\sqrt{3})$. We can work out the bases on these double cones from (\ref{offset}) and Table \ref{tab0}, and the results are given in Table \ref{tab1}. We see that the basis on $\theta_{b}$ is its own inverse in the origin (which we already knew from Proposition 2) while the two bases on $\theta_{a}$ are the inverses of each other. Thus the MUB as a whole is inversion symmetric and not chiral. We call this MUB MUB-1 to distinguish it from MUB-2 that we will introduce shortly. In addition to listing the M-vectors of this MUB, we have also listed the corresponding rays in Hilbert space (see Appendix 3 for an explanation of how the rays can be worked out). The last column shows the rays in the simplified form in which they are sometimes quoted; if we denote by $Z$, $X$, $Y=XZ$ and $W=XZ^{2}$ the operators of the qutrit Pauli group, then the bases in the four blocks are the simultaneous eigenstates of the pairs of commuting operators ($Z,Z^{2}$), ($Y,Y^{2}$), ($W,W^{2}$) and ($X,X^{2}$), respectively.\\

\noindent
Consider next the MUB defined by $y=0$, for which the quartic reduces to

\begin{equation}\label{caseB}
(x^2+10x-3)^2=0 \hspace{2mm}.
\end{equation}

\noindent
The solution again consists of a pair of double roots, $2\sqrt{7}-5$ and $-2\sqrt{7}-5$, of which only the former is acceptable and leads to the double-cone $\theta_{d}=\cos^{-1}(\sqrt{2\sqrt{7}-5})$. The other double-cone, corresponding to $y=0$, has angle $\pi/2$ and is actually the x-y plane. We can work out the bases on these double-cones from (\ref{offset}) and Table \ref{tab0} and the results are given in Table \ref{tab1}. One sees that the double-cone $\theta_{d}$ has two bases that are the reflections of each other in the horizontal plane while the flattened double-cone has a single basis that is its own reflection. Thus the MUB as a whole has reflection symmetry is not chiral. The rays of this MUB are also listed and, again, a simple unitary transformation can be used to turn them into the simple form shown in the last column.\\

\noindent
Finally we address an important point. We have shown how to construct an infinite number of MUBs by starting from an arbitrary state on the double-cone $\theta=\cos^{-1}\sqrt{y}$, for any $0\leq y \leq \frac{1}{3}$. One can ask if these MUBs are independent or related to each other. The answer is that they are all unitarily equivalent to each other under a  ``phasing'' transformation that we now describe. Consider the rays of MUB-1, shown in last column of Table 1. Applying the unitary transformation $U = \text{diag}(1,1,e^{i\psi})$ to these rays, with $0\leq \psi < 2\pi$, allows us to generate all the MUBs yielded by our construction. For $\psi=0, \frac{2\pi}{3}$ or $\frac{4\pi}{3}$ we get MUB-1, which has inversion symmetry; for $\psi=\pi$ we get MUB-2, which has reflection symmetry; and for all other $\psi$ we get MUBs that are chiral. The MUBs for $\psi=\pi-\psi_{0}$ and $\psi=\pi+\psi_{0}$ are the inverses of each other in the origin (up to a rotation about the z-axis) for all $\psi_{0}$ in the open interval $0< \psi_{0}<\pi$, with the exception of the point $\frac{2\pi}{3}$. In conclusion a spin-1 system has a unique MUB, although it can assume an infinite number of guises when viewed in the Majorana representation.\\

\begin{table}[ht]
\centering 
\begin{tabular}{|c | c | c | c |} 
\hline 
   &   &   &  \\
 State  &  $(\theta_{1},\phi_{1}|\theta_{2},\phi_{2})$ & Ray & Simplified  \\
    &   &   &  \\
\hline
   &   &   &  \\
 C & $(0,0|0,0) $ &  $(1,0,0)$  & $(1,0,0)$ \\
   &   &   &  \\
 A &$(0,0|\pi,0)$ &  $(0,1,0)$  & $(0,1,0) $ \\
   &   &   &  \\
 C &$(\pi,0|\pi,0)$ &  $(0,0,1)$  & $(0,0,1) $ \\
   &   &   &  \\
\hline
   &   &   &  \\
$D^{R}(\theta_{a},0)$  & $(\theta_{a},0|\pi-\theta_{a},\pi-\phi_{a})$  &  $(1,c\omega,-c^{2})$  & $(1,1,\omega)$ \\
   &   &   &  \\
$D^{R}(\theta_{a},\frac{2\pi}{3})$ & $(\theta_{a},\frac{2\pi}{3}|\pi-\theta_{a},\frac{5\pi}{3}-\phi_{a})$ &  $(1,c\omega^{2},-c^{2}\omega^{2})$  & $(1,\omega,1) $ \\
   &   &   &  \\
$D^{R}(\theta_{a},\frac{4\pi}{3})$  & $(\theta_{a},\frac{4\pi}{3}|\pi-\theta_{a},\frac{\pi}{3}-\phi_{a})$ &
  $(1,c,-c^{2}\omega)$  & $(1,\omega^{2},\omega^{2}) $ \\
  &   &   &  \\
\hline
  &   &   &  \\
$D^{L}(\theta_{a},-\phi_{a})$  & $(\theta_{a},-\phi_{a}|\pi-\theta_{a},\pi)$  & $(1,c\omega^{2},-c^{2})$  & $(1,\omega,\omega) $  \\
   &   &   &  \\
$D^{L}(\theta_{a},\frac{2\pi}{3}-\phi_{a})$  & $(\theta_{a},\frac{2\pi}{3}-\phi_{a}|\pi-\theta_{a},\frac{5\pi}{3})$  &  $(1,c,-c^{2}\omega^{2})$  & $(1,\omega^{2},1) $ \\
   &   &   &  \\
$D^{L}(\theta_{a},\frac{4\pi}{3}-\phi_{a})$  & $(\theta_{a},\frac{4\pi}{3}-\phi_{a}|\pi-\theta_{a},\frac{\pi}{3})$ &  $(1,c\omega,-c^{2}\omega)$  & $(1,1,\omega^{2}) $\\
&   &   &  \\
\hline
 &   &   &  \\
$D(\theta_{b},\pi-\frac{\phi_{a}}{2})$   &$(\theta_{b},\pi-\frac{\phi_{a}}{2}|\pi-\theta_{b},-\frac{\phi_{a}}{2})$  &  $(1,c,-c^{2})$  & $(1,\omega^{2},\omega)$ \\
   &   &   &  \\
  $D(\theta_{b},\frac{5\pi}{3}-\frac{\phi_{a}}{2})$ & $(\theta_{b},\frac{5\pi}{3}-\frac{\phi_{a}}{2}|\pi-\theta_{b},\frac{2\pi}{3}-\frac{\phi_{a}}{2})$  &  $(1,c\omega,-c^{2}\omega^{2})$  & $(1,1,1) $ \\
   &   &   &  \\
$D(\theta_{b},\frac{\pi}{3}-\frac{\phi_{a}}{2})$  & $(\theta_{b},\frac{\pi}{3}-\frac{\phi_{a}}{2}|\pi-\theta_{b},\frac{4\pi}{3}-\frac{\phi_{a}}{2})$  &  $(1,c\omega^{2},-c^{2}\omega)$  & $(1,\omega,\omega^{2}) $ \\
&   &   &  \\
\hline
\end{tabular}
\caption{MUB-1 of a spin-1 system, with the states of each basis shown in a separate block. The first column lists the states as D-states and the second gives the angular parameters of their M-vectors, with $\theta_{a}=\cos^{-1}(\sqrt{\frac{2}{\sqrt{3}}-1}) = 66.84^{\circ}$ , $\theta_{b} = \cos^{-1}(1/\sqrt{3})=54.74^{\circ}$ and $\phi_{a} =\cos^{-1}(\frac{\sqrt{3}-1}{2}) = 68.53^{\circ}$. The third column shows the states as rays in $\mathbb{CP}^{2}$, with $\omega = \exp{(2\pi i/3)}$ and $c = \exp({-i\phi_{a}/2})$, and the fourth column shows the rays in a simplified form obtained by multiplying the vectors in the third column by the unitary matrix $U = \text{diag}(1,\omega^{2}c^{-1},-\omega c^{-2})$. The M-vectors of this MUB have two geometrical symmetries: a threefold rotational symmetry about the z-axis (with the states of the first basis being invariant and those of the other three cycling into each other) and inversion symmetry in the origin (with the first and last bases being invariant and the other two going into each other).}
\label{tab1} 
\end{table}

\begin{table}[ht]
\centering 
\begin{tabular}{|c | c | c | c |} 
\hline 
   &   &   &  \\
 State  &  $(\theta_{1},\phi_{1}|\theta_{2},\phi_{2})$ & Ray & Simplified  \\
    &   &   &  \\
\hline
   &   &   &  \\
 C & $(0,0|0,0) $ &  $(1,0,0)$  & $(1,0,0)$ \\
   &   &   &  \\
 A &$(0,0|\pi,0)$ &  $(0,1,0)$  & $(0,1,0) $ \\
   &   &   &  \\
 C &$(\pi,0|\pi,0)$ &  $(0,0,1)$  & $(0,0,1) $ \\
   &   &   &  \\
\hline
   &   &   &  \\
$D^{R}(\theta_{d},0)$  & $(\theta_{d},0|\pi-\theta_{d},\pi-\phi_{d})$  &  $(1,-ik\omega^{2},-k^{2})$  & $(1,\omega,\omega)$ \\
   &   &   &  \\
$D^{R}(\theta_{d},\frac{2\pi}{3})$ & $(\theta_{d},\frac{2\pi}{3}|\pi-\theta_{d},\frac{5\pi}{3}-\phi_{d})$ &  $(1,-ik,-k^{2}\omega^{2}$  & $(1,\omega^{2},1) $ \\
   &   &   &  \\
$D^{R}(\theta_{d},\frac{4\pi}{3})$  & $(\theta_{d},\frac{4\pi}{3}|\pi-\theta_{d},\frac{\pi}{3}-\phi_{d})$ &
  $(1,-ik\omega,-k^{2}\omega)$  & $(1,1,\omega^{2}) $ \\
  &   &   &  \\
\hline
  &   &   &  \\
$D^{L}(\theta_{d},\pi-\phi_{d})$  & $(\theta_{d},\pi-\phi_{d}|\pi-\theta_{d},0)$  & $(1,-ik\omega,-k^{2})$  & $(1,1,\omega) $  \\
   &   &   &  \\
$D^{L}(\theta_{d},\frac{5\pi}{3}-\phi_{d})$  & $(\theta_{d},\frac{5\pi}{3}-\phi_{d}|\pi-\theta_{d},\frac{2\pi}{3})$  &  $(1,-ik\omega^{2},-k^{2}\omega^{2})$  & $(1,\omega,1) $ \\
   &   &   &  \\
$D^{L}(\theta_{d},\frac{\pi}{3}-\phi_{d})$  & $(\theta_{d},\frac{\pi}{3}-\phi_{d}|\pi-\theta_{d},\frac{4\pi}{3})$ &  $(1,-ik,-k^{2}\omega)$  & $(1,\omega^{2},\omega^{2}) $\\
&   &   &  \\
\hline
 &   &   &  \\
$D(\frac{\pi}{2},\frac{5\pi}{4}-\frac{\phi_{d}}{2})$   &$(\frac{\pi}{2},\frac{5\pi}{4}-\frac{\phi_{d}}{2}|\frac{\pi}{2},\frac{7\pi}{4}-\frac{\phi_{d}}{2})$  &  $(1,-ik,-k^{2})$  & $(1,\omega^{2},\omega)$ \\
   &   &   &  \\
$D(\frac{\pi}{2},\frac{23\pi}{12}-\frac{\phi_{d}}{2})$   &$(\frac{\pi}{2},\frac{23\pi}{12}-\frac{\phi_{d}}{2}|\frac{\pi}{2},\frac{29\pi}{12}-\frac{\phi_{d}}{2})$  &  $(1,-ik,-k^{2}\omega^{2})$  & $(1,1,1)$ \\
   &   &   &  \\
$D(\frac{\pi}{2},\frac{31\pi}{12}-\frac{\phi_{d}}{2})$   &$(\frac{\pi}{2},\frac{31\pi}{12}-\frac{\phi_{d}}{2}|\frac{\pi}{2},\frac{37\pi}{12}-\frac{\phi_{d}}{2})$  &  $(1,-ik\omega^{2},-k^{2}\omega)$  & $(1,\omega,\omega^{2})$ \\
&   &   &  \\
\hline
\end{tabular}
\caption{MUB-2 of a spin-1 system, in a format similar to that of Table 1, with the angles $\theta_{d}=\cos^{-1}(\sqrt{2\sqrt{7}-5})=57.32^{\circ}$ and $\phi_{d}=\cos^{-1}(\frac{\sqrt{7}-1}{2})=34.63^{\circ}$. The third column shows the states as rays in $\mathbb{CP}^{2}$, with $\omega = \exp{(2\pi i/3)}$ and $k = \exp({-i\phi_{d}/2})$. The fourth column shows the rays in a simplified form, obtained by multiplying the rays in the third column by the unitary matrix $U = \text{diag}(1,i\omega^{2}k^{-1},-\omega k^{-2})$. The M-vectors of this MUB have two geometrical symmetries: a threefold rotational symmetry about the z-axis (with the states of the first basis being invariant and those of the other three cycling into each other) and reflection symmetry in the horizontal plane (with the first and last bases being invariant and the other two going into each other).}
\label{tab2} 
\end{table}

\section{\label{sec:4} SICs of a spin-1 system}

A SIC of a spin-1 system is a set of nine (normalized) states with the property that the overlap of any two of them is $1/4$. If two states are members of a SIC, then their M-vectors must obey the equiangularity condition (\ref{equi}). Let us first try to identify two states of the same type (i.e., C,A or D) that are equiangular. Begin with the case of C-states. Putting $\vec{a_{1}} = \vec{a_{2}} =\vec{a}$ and $\vec{b_{1}} = \vec{b_{2}} =\vec{b}$ in (\ref{equi}) gives $(\vec{a}\cdot\vec{b})\big[(\vec{a}\cdot\vec{b})+2\big] = 0$, whose only solution is $\vec{a}\cdot\vec{b} = 0$. This implies that two C-states are equiangular only if their axes are orthogonal. Three C-states whose axes are mutually orthogonal are also mutually equiangular, and it is clear that this is the maximum number of mutually equiangular C-states that there can be. Note that this triad of C-states has a threefold axis of symmetry that is equally inclined to their M-vectors.\\

\noindent
Let us next try to identify a pair of A-states that are equiangular. Putting $\vec{a_{1}} = -\vec{a_{2}} =\vec{a}$ and $\vec{b_{1}} = -\vec{b_{2}} =\vec{b}$ in (\ref{equi}) gives $\vec{a}\cdot\vec{b} = \pm 1/2$, which implies that two A-states are equiangular only if their axes intersect at an angle of $60^{\circ}$ (or $120^{\circ}$). It is not hard to see that one can find a third A-state equiangular to the previous two and that it can be done in two ways: the states can either lie in a plane with their axes pointing towards the vertices of a regular hexagon, or else they can be equally spaced around the surface a double-cone of semi-vertex angle $\cos^{-1}(\sqrt{2/3})$. Both these configurations have a threefold axis of symmetry, like the triad of C-states found above.\\

\noindent
These observations lead us to define an \textit{equiangular triad} (or etriad) as a set of three mutually equiangular states whose M-vectors are related by a threefold rotation about a common axis. Taking the rotation axis as the z-axis, we can write the M-vectors of the states (in angular form) as
\begin{equation}
        (\theta_{1},\phi_{1}|\theta_{2},\phi_{2})  \hspace{0.5mm} , \hspace{0.5mm} (\theta_{1},\phi_{1}+\frac{2\pi}{3}|\theta_{2},\phi_{2}+\frac{2\pi}{3}) \hspace{1mm} , \hspace{1mm}
        (\theta_{1},\phi_{1}+\frac{4\pi}{3}|\theta_{2},\phi_{2}+\frac{4\pi}{3}) \hspace{0.5mm}.
\label{triadgen}
\end{equation}

\noindent
The general equation determining all possible etriads can be found by substituting any pair of these states into (\ref{equi}), whereupon one gets

\noindent
\begin{multline}
\sin^{2}\theta_{1}\sin^{2}\theta_{2}\cos^{2}(\phi_{2}-\phi_{1})-2\sin\theta_{1}\sin\theta_{2}(1+3\cos\theta_{1}\cos\theta_{2})\cos(\phi_{2}-\phi_{1})
\\ -3 +4\cos^{2}\theta_{1}+4\cos^{2}\theta_{2}+6\cos\theta_{1}\cos\theta_{2})+5\cos^{2}\theta_{1}\cos^{2}\theta_{2} = 0 \hspace{0.5mm}.
\label{etriadgen}
\end{multline}

\noindent
We can pick out a number of interesting etriads from (\ref{etriadgen}) by specializing it in various ways:
\\

\noindent
(a) If we put $\theta_{1}=\theta_{2}$ and $\phi_{1}=\phi_{2}$ we find that $\theta_{1}=\theta_{2}=\cos^{-1}(1/\sqrt{3})$. This is just the C-state etriad we found earlier.
\\

\noindent
(b) Putting $\theta_{2}=\pi-\theta_{1}$ and $\phi_{2}=\phi_{1}+\pi$ gives the two solutions $\theta_{1}=\pi/2$ and $\theta_{1}=\cos^{-1}\sqrt{2/3}$. These describe the two A-state etriads we found earlier.
\\

\noindent
(c) Next we consider etriads made up of D-states. If we put $\theta_{1}=\phi_{1}=0$ we find that $\theta_{2}=\cos^{-1}(-1/3)$, with $\phi_{2}$ being arbitrary. The M-vectors of the D-states point towards the vertices of a tetrahedron, with the vector along the z-axis being common to all the states and one each of the remaining vectors going with each state.
\\

\noindent
(d) A second etriad involving D-states can be obtained by putting $\theta_{1}=\theta_{2}= \cos^{-1}(1/3)$, whereupon one finds that $\phi_{2}=\phi_{1}+2\pi/3$. The D-states are then described by three equally spaced vectors around the surface of a cone, with two vectors going with each of the states and any two states having a vector in common.
\\

\noindent
The details of these etriads are summarized in Table \ref{tabetriad} and Figs. 6 and 7 show them in pictorial form.
\\

\begin{table}[ht]
\centering 
\begin{tabular}{|c | c | c |} 
\hline 
   &   &   \\
 Label & First Member & Geometry  \\
    &   &  \\
\hline
   &   &   \\
  C & $\Big(\cos^{-1}(\frac{1}{3}),0\Big|\cos^{-1}(\frac{1}{3}),0\Big)$  & C-states along the edges of a cube  \\
     &   &   \\
\hline
   &   &   \\
 A1 & $\Big(\frac{\pi}{2},0\Big|\frac{\pi}{2},\pi\Big)$ & A-states in a plane      \\
   &   &   \\
\hline
   &   &   \\
 A2 & $\Big(\cos^{-1}(\sqrt{\frac{2}{3}}),0\Big|\pi-\cos^{-1}(\sqrt{\frac{2}{3}}),\pi\Big)$ & A-states on a double-cone      \\
   &   &    \\
\hline
   &   &   \\
D1 & $\Big(0,0\Big|\cos^{-1}(-\frac{1}{3}),\phi_{2}\Big)$ & Tetrahedral configuration    \\
   &   &    \\
\hline
   &   &   \\
D2 & $\Big(\cos^{-1}(\frac{1}{3}),0\Big|\cos^{-1}(\frac{1}{3}),\frac{2\pi}{3}\Big)$ & Three vectors equally spaced around a cone  \\
   &   &    \\
\hline
\end{tabular}
\caption{Some etriads (i.e., sets of three mutually equiangular states). Only the first member of each etriad is shown (the second and third members can be obtained as indicated in Eq.(\ref{triadgen})). The last column briefly describes the geometry of the M-vectors in each etriad. See Figs. 6(a),6(b),7(a) and 7(b) for depictions of A1, A2, D1 and D2.}
\label{tabetriad} 
\end{table}

\begin{figure}[htp]
\begin{center}
\includegraphics[width=.80\textwidth]{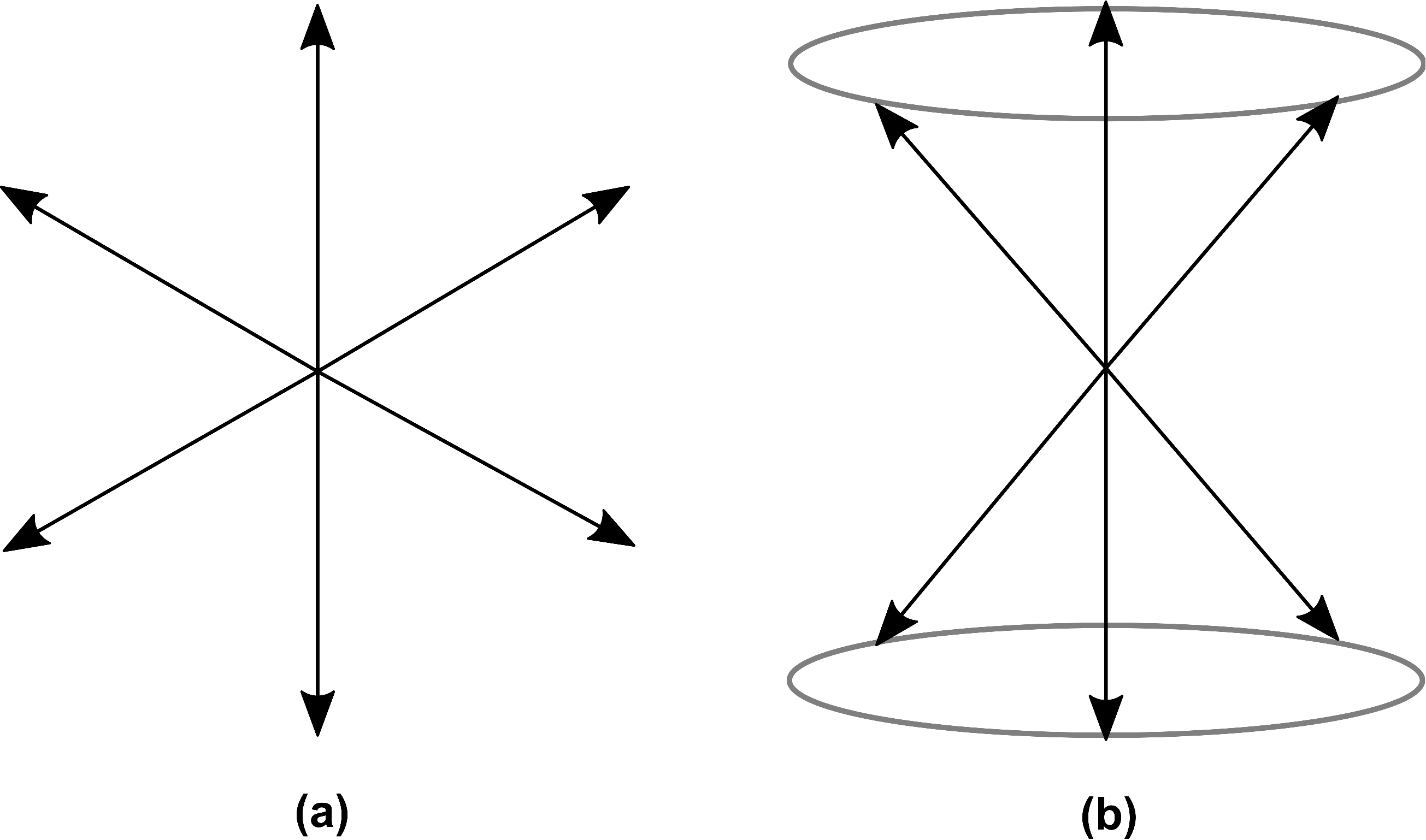}
\end{center}
\caption{(a) The etriad A1, whose three A-states have M-vectors lying in a plane and pointing towards the vertices of a regular hexagon. (b) The etriad A2, whose A-states have M-vectors lying on the surface of a double-cone of semi-vertex angle $\cos^{-1}\sqrt{2/3}$ at equally spaced intervals.}
\label{Fig.6}
\end{figure}

\begin{figure}[htp]
\begin{center}
\includegraphics[width=.80\textwidth]{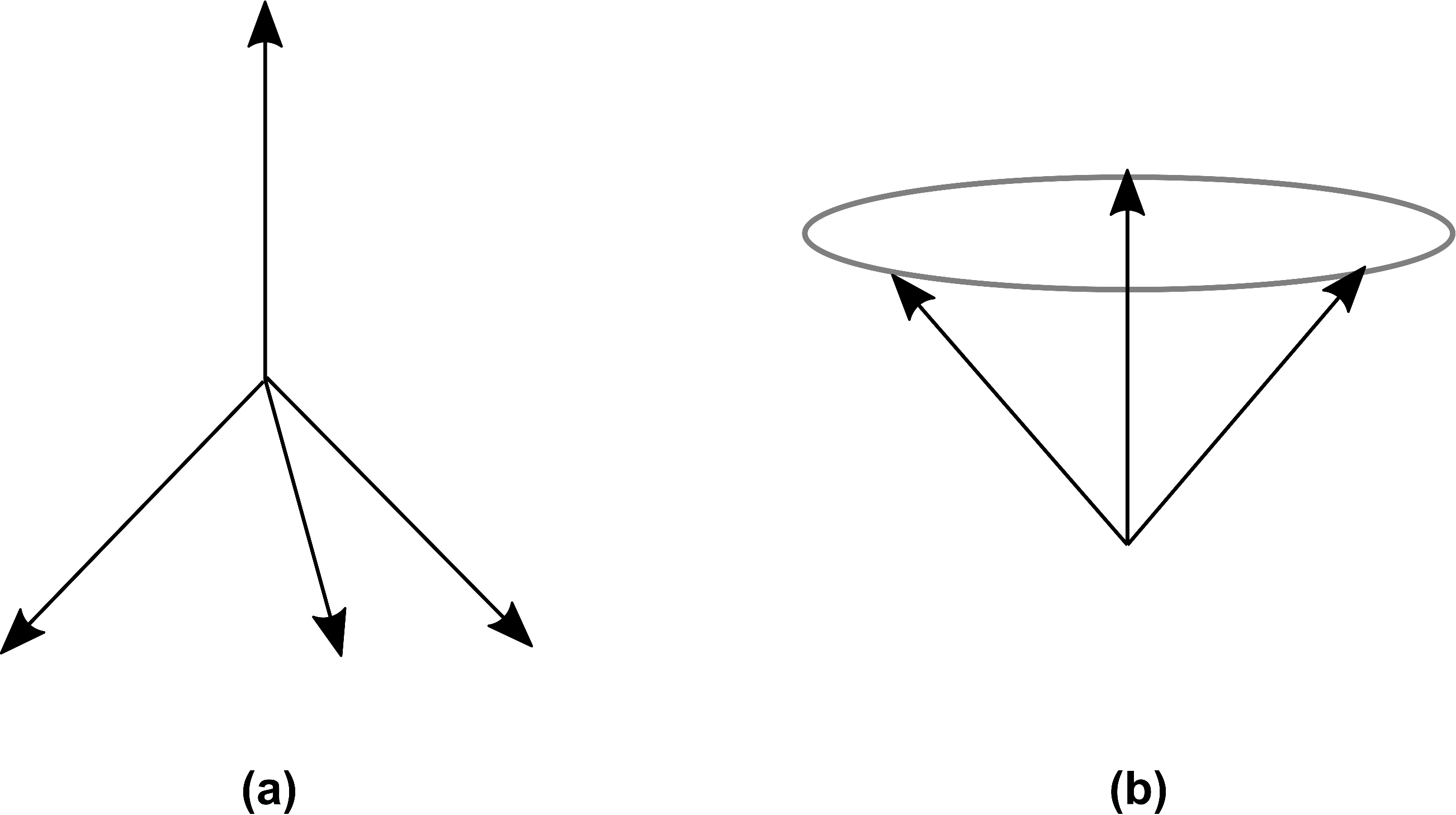}
\end{center}
\caption{(a) The etriad D1, whose states have M-vectors pointing towards the vertices of a regular tetrahedron; the upward vector belongs to all three states and each downward vector to one of the states. Inverting these vectors in the origin gives the M-vectors of the etriad D1$^{-1}$. Uniting the etriads D1 and D1$^{-1}$ with A1 gives the M-vectors of SIC-1, which are described in detail in Table \ref{tabSIC1}. (b) The etriad D2, whose states are described by three M-vectors spaced equally around the surface of a cone of semi-vertex angle $\arccos(1/3)$; each state is described by a pair of the vectors and any two states have a vector in common. Inverting these vectors in the origin gives the M-vectors of the etriad D2$^{-1}$. Uniting the etriads D2 and D2$^{-1}$ with A2 gives the M-vectors of SIC-2, which are described in detail in Table \ref{tabSIC2}.}
\label{Fig.7}
\end{figure}

\noindent
The problem now is to see whether a given etriad can be extended into a SIC. We do this by trying to find a D-state that is equiangular to each of the states of a given etriad. This leads, via (\ref{equi}), to three equations that must be satisfied by the parameters of the sought for D-state. If the initial etriad is a propitious one, these equations will have multiple solutions that yield all the additional states needed to complete a SIC. We find that if we start with either of the etriads A1 or A2, the equations can be solved and lead to six additional states that complete a SIC. In the case of A1, the remaining states are those of the etriad D1 and the etriad D1$^{-1}$ obtained by inverting all the M-vectors of D1 in the origin:
\begin{equation}
  \text{SIC-1 = A1 + D1 + D1$^{-1}$}
\label{SIC1a}
\end{equation}

\noindent
The SIC arising from A2 can be written similarly as
\begin{equation}
  \text{SIC-2 = A2 + D2 + D2$^{-1}$} \hspace{2mm},
\label{SIC2a}
\end{equation}

\noindent
where again D2$^{-1}$ is the etriad obtained by inverting all the M-vectors of D2 in the origin. We have termed these SICs SIC-1 and SIC-2 for convenience. Both SICs consist of etriads with a common symmetry axis, so they each have threefold axis as a whole.
\\

\noindent
An interesting feature of SIC-1 is that its component etriads can be rotated independently about the z-axis without destroying the SIC. The SIC is shown in this more general form in Table \ref{tabSIC1}, rather than in the restricted form of Eq.(\ref{SIC1a}). This SIC is essentially the same as that given in  \cite{appleby1}.
\\

\begin{table}[ht]
\centering 
\begin{tabular}{|c | c | c | c |} 
\hline 
   &   &     \\
 State  &  $(\theta_{1},\phi_{1}|\theta_{2},\phi_{2})$ & Ray  \\
    &   &    \\
\hline
 &   &    \\
 1 & $(\frac{\pi}{2},0|\frac{\pi}{2},\pi)$  &  $(1,0,-1)$  \\
   &   &     \\
 2 & $(\frac{\pi}{2},\frac{\pi}{3}|\frac{\pi}{2},\frac{4\pi}{3})$  &  $(1,0,-\omega)$  \\
   &   &     \\
 3  & $(\frac{\pi}{2},\frac{2\pi}{3}|\frac{\pi}{2},\frac{5\pi}{3})$  &  $(1,0,-\omega^{2})$  \\
   &   &     \\
\hline
  &   &     \\
4 & $(0,0|\pi-\theta_{0},\phi_{a})$  &  $(1,e^{i\phi_{a}},0)$  \\
   &   &     \\
5 & $(0,0|\pi-\theta_{0},\phi_{a}+\frac{2\pi}{3})$  &  $(1,\omega e^{i\phi_{a}},0)$  \\
   &   &     \\
6 & $(0,0|\pi-\theta_{0},\phi_{a}+\frac{4\pi}{3})$  &  $(1,\omega^{2} e^{i\phi_{a}},0)$  \\
   &   &     \\
\hline
 &   &     \\
7 & $(\pi,0|\theta_{0},\phi_{b})$  &  $(0,1,e^{i\phi_{b}})$  \\
   &   &     \\
8 & $(\pi,0|\theta_{0},\phi_{b}+\frac{2\pi}{3})$  &  $(0,1,\omega e^{i\phi_{b}})$  \\
   &   &     \\
9 & $(\pi,0|\theta_{0},\phi_{b}+\frac{4\pi}{3})$  &  $(0,1,\omega^{2} e^{i\phi_{b}})$  \\
   &   &     \\
\hline
\end{tabular}
\caption{SIC-1 of a spin-1 system, with each of the three etriads shown in a separate block. The second column shows the angular parameters of the M-vectors of the states and the third shows the states as rays in $\mathbb{CP}^{2}$ (note: $\theta_{0}=\arccos(-\frac{1}{3})$, $\omega = \exp{(2\pi i/3)}$ and the phases $\phi_{a}$ and $\phi_{b}$ are arbitrary). The states 1,2,3 make up the etriad A1, while 4,5,6 and 7,8,9 make up the etriads D1 and D1$^{-1}$, respectively, but with the arbitrary phases $\phi_{a}$ and $\phi_{b}$ added.}
\label{tabSIC1} 
\end{table}

\noindent
SIC-2, shown in Table \ref{tabSIC2}, consists of three etriads that are locked rigidly relative to each other. This SIC is unitarily equivalent to the Norrell states discussed by Veitch et al \cite{veitch} and Stacey \cite{stacey2}. The Norrell states are pure states of a spin-1 particle with the property that two elements of their discrete Wigner function are equal to $-\frac{1}{6}$. There are 36 such states altogether, and they fall into 4 SICs of 9 states each. One of these SICs is unitarily equivalent to SIC-2\footnote{The precise connection is as follows. Begin with the rays of SIC-2, given in the last column of Table \ref{tabSIC2}, and change the signs of all the third components (which amounts to a unitary transformation). Then construct the phase space point operators $A_{j}$ in terms of the Heisenberg-Weyl operators, as explained in \cite{veitch} and \cite{stacey2}, and use them to calculate the Wigner function elements of the SIC-2 states via the relation $W_{j}=\frac{1}{9}\text{Tr}(\rho A_{j})$, where $\rho$ is the density matrix of any of the states; two of the Wigner function elements will be found to be $-\frac{1}{6}$, six will be found to be $\frac{1}{6}$ and one $\frac{1}{3}$, which is just the signature of a Norrell state.}.
\\

\begin{table}[ht]
\centering 
\begin{tabular}{|c | c | c | c |} 
\hline 
   &   &     \\
 State  &  $(\theta_{1},\phi_{1}|\theta_{2},\phi_{2})$ & Ray  \\
    &   &    \\
\hline
 &   &    \\
 1 &  $(\theta_{f},0|\pi-\theta_{f},\pi)$ & $(1,-2,-1)$  \\
   &   &     \\
 2 &  $(\theta_{f},\frac{2\pi}{3}|\pi-\theta_{f},\frac{5\pi}{3})$ & $(1,-2\omega,-\omega^{2})$  \\
   &   &     \\
 3 &  $(\theta_{f},\frac{4\pi}{3}|\pi-\theta_{f},\frac{\pi}{3})$ & $(1,-2\omega^{2},-\omega)$  \\
   &   &     \\
\hline
  &   &     \\
4 & $(\theta_{g},0|\theta_{g},\frac{2\pi}{3})$  & $(2,-\omega^{2},\omega)$  \\
   &   &     \\
5 &  $(\theta_{g},\frac{2\pi}{3}|\theta_{g},\frac{4\pi}{3})$ & $(2,-1,1)$  \\
   &   &     \\
6 & $(\theta_{g},\frac{4\pi}{3}|\theta_{g},0)$  & $(2,-\omega,\omega^{2})$  \\
   &   &     \\
\hline
 &   &     \\
7 & $(\pi-\theta_{g},\pi|\pi-\theta_{g},\frac{5\pi}{3})$  &  $(1,\omega^{2},2\omega)$  \\
   &   &     \\
8 & $(\pi-\theta_{g},\frac{5\pi}{3}|\pi-\theta_{g},\frac{\pi}{3})$  &  $(1,1,2)$  \\
   &   &     \\
9 & $(\pi-\theta_{g},\frac{\pi}{3}|\pi-\theta_{g},\pi)$  &  $(1,\omega,2\omega^{2})$  \\
   &   &     \\
\hline
\end{tabular}
\caption{SIC-2 of a spin-1 system, with each of the three etriads shown in a separate block. The second column shows the angular parameters of the M-vectors of the states and the third shows the states as rays in $\mathbb{CP}^{2}$ (note: $\theta_{f}=\arccos(\sqrt{2/3})$, $\theta_{g}=\arccos(1/3)$ and $\omega = \exp{(2\pi i/3)}$). The states 1,2,3 make up the etriad A2, while 4,5,6 and 7,8,9 make up the etriads D2 and D2$^{-1}$, respectively. The rays in the last column are identical to the Norrell states if the signs of their third components are reversed.}
\label{tabSIC2} 
\end{table}

\pagebreak

\section{\label{sec:5} Discussion}

\noindent
This paper shows how the MUBs and SICs of a spin-1 system can be derived from the formula (2) for the overlap of two states in terms of their Majorana vectors. The formula allows the notions of orthogonality, unbiasedness and equiangularity, which are usually defined for vectors in Hilbert space, to be translated into relations between  Majorana vectors and used, in conjunction with simple geometrical arguments, to deduce the MUBs and the SICs. The result is a view of the MUBs and SICs as symmetrical collections of vectors in ordinary three-dimensional space, rather than as rays in an abstract (and complex) Hilbert space. The message of this paper is that if one is willing to start from the formula (2), one can work out the MUBs and SICs in ordinary three-dimensional space without making a detour into Hilbert space.\\

\noindent
It should remarked that the overlap formula (2) contains only information about the magnitude of the inner product and not its phase. However the phase information is not needed in the derivation of the MUBs or the SICs.\\

\noindent
Let us briefly recapitulate the view of the MUBs and the SICs given by the Majorana approach.\\

\noindent
Consider first the MUBs. Recall that any state is represented by a pair of unit vectors, that we refer to as its Majorana vectors or M-vectors. If we choose as the first basis of a MUB the standard angular momentum basis, all of whose M-vectors point up or down along the z-axis, then all the remaining states are forced to lie on a set of nested double-cones about the z-axis. One can pick an arbitrary state on an arbitrary double-cone as a seed for a MUB, and then all its other states are automatically determined. The generic MUB consists of a central axis, containing the angular momentum basis, and three double-cones, each accommodating a basis. It is chiral, and inverting all its M-vectors in the origin gives rise to another MUB of the opposite handedness. However there are two special MUBs for which the three double-cones collapse into two, with one accommodating a pair of bases and the other just one. One of these MUBs has inversion symmetry in the origin and the other reflection symmetry in the horizontal plane, so neither is chiral. All the MUBs, both the chiral and non-chiral ones, are part of a single family whose members can be made to pass into each other by a one-parameter family of unitary transformations (discussed at the end of Sec. \ref{sec:3}).\\

\noindent
Turning next to the SICs, the distinctive feature of the Majorana approach is that it allows any SIC to be built up as a triad of etriads, where each etriad is a set of three mutually equiangular states whose Majorana vectors are related by a threefold rotation about a central axis. Although there is a general equation (\ref{triadgen}) that determines all possible etriads, there is unfortunately no way of telling which etriads can actually be extended into SICs, making this approach difficult to exploit. However we did identify two simple etriads, both made up of A-states, that could be extended into SICs. One of them, which we refered to as SIC-1, has the interesting feature that its constituent etriads can be rotated arbitrarily relative to each (about their common axis) without destroying the SIC; it is well known \cite{appleby1} and includes the Hesse SIC\footnote{It is worth noting that the Hesse SIC can be united with a MUB of a spin-1 system to yield a system of 21 rays, the so-called Hesse configuration \cite{hesse}, that can be used to derive a state-independent inequality that is satisfied by any noncontextual hidden variables theory but is violated by quantum mechanics \cite{beng}.} among its members. The other, which we referred to as SIC-2, is equivalent to the Norrell SIC which is of interest as one of the maximally magic resources for quantum computation \cite{veitch},\cite{stacey2}. \\

\noindent
However it is well known \cite{caves2004}, \cite{hughston}, \cite{zhu2} that a spin-1 system has an infinite number of  SICs, and the Majorana approach does not seem to be capable of ferreting them out. Whether this is due to the inherent limitations of the approach, or my ineptness in milking it, I am unable to tell. As far as a geometrical construction of SICs is concerned, the work of Hughston and Salamon \cite{hughston} provides probably the most satisfying and rewarding approach. These authors use a moment mapping between $\mathbb{CP}^{2}$ and $\mathbb{R}^{3}$ to cast many features of the problem in familiar geometrical terms and show, by a predominantly analytic method (aided by a computation at the end), that all SICs are unitarily equivalent to one of a particularly simple kind they term a midpoint solution.\\

\noindent
To sum up, the Majorana approach gives a complete account of the spin-1 MUBs but seems to be unable to do the same for the SICs. I would be delighted if someone were to prove me to be unduly pessimistic.
\\

{\bf Acknowledgements}. I would like to thank Tenzin Kalden, Eric Reich and Junjiang Le, all undergraduate students at WPI, for their participation in different phases of this project and for their help in checking some of the results presented here. I thank Professor Bill Martin of the Mathematics Department at WPI for discussions on the mathematical aspects of SICs. I am also very grateful to Dr. Marcus Appleby for a long meeting during which he shared his insights on the matters discussed here, answered many questions and also provided me with a host of useful references. Finally I would like to thank two anonymous referees for their careful reading of an earlier draft of the paper and for pointing out some erroneous claims I made. They also suggested some improvements to the presentation and supplied me with some valuable references I had missed. Needless to say, any shortcomings that remain in this work are entirely mine.
\\

\pagebreak

\appendix
\section{\label{App1} Appendix 1: Derivation of Overlap formula for spin-1 states}

A pure state of a spin-half particle can be represented by a point on the unit sphere and written $|\vec{a} \rangle$, where $\vec{a}$ is the unit vector from the center of the sphere to the representative point and it is assumed that the state is normalized, i.e. $\langle\vec{a}|\vec{a}\rangle=1$. In the Majorana approach, an arbitrary pure state of a spin-1 particle is written as the symmetrized outer product of the states of a pair of fictitious spin-half particles,

\begin{equation}
|\vec{a_{1}},\vec{a_{2}}\rangle =\frac{|\vec{a_{1}}\rangle_{1}\otimes|\vec{a_{2}}\rangle_{2} + |\vec{a_{2}}\rangle_{1}\otimes|\vec{a_{1}}\rangle_{2}}{\sqrt{3+\vec{a_{1}}\cdot\vec{a_{2}}}} \hspace{2mm} ,
\label{A1}
\end{equation}

\noindent
where the subscripts on the kets refer to the particles while those on the vectors are merely state labels, and the denominator is a normalization factor that follows from the fact that $_{i}\langle \vec{a}|\vec{a}\rangle_{j} =\delta_{ij}$ and $_{i}\langle \vec{a}|\vec{b}\rangle_{j} = \frac{1}{2}( 1+\vec{a}\cdot\vec{b})\delta_{ij}$, where $i,j=1,2$. We wish to calculate the overlap of the state (\ref{A1}) with the state $|\vec{b_{1}},\vec{b_{2}}\rangle$, i.e., the quantity $|\langle \vec{b_{1}},\vec{b_{2}}|\vec{a_{1}},\vec{a_{2}}\rangle|^{2}$. A quick way of getting the answer is to note, from (\ref{A1}), that the numerator of the overlap must have the following properties: (a) it must be a function of all the scalar products that can be constructed out of the vectors $\vec{a_{1}},\vec{a_{2}},\vec{b_{1}}$ and $\vec{b_{2}}$, (b) it must have terms that are both linear and quadratic in these scalar products, and (c) it must be invariant under an interchange of $\vec{a_{1}}$ and $\vec{a_{2}}$ or $\vec{b_{1}}$ and $\vec{b_{2}}$ (or both of these operations performed together). An expression for the overlap having all these properties is

\begin{equation}
|\langle \vec{b_{1}},\vec{b_{2}}|\vec{a_{1}},\vec{a_{2}}\rangle|^{2} = \frac{x_{1}+x_{2}L_{1}+x_{3}L_{2}+x_{4}Q_{1}+x_{5}Q_{2}}{(3+\vec{a_{1}}\cdot\vec{a_{2}}) (3+\vec{b_{1}}\cdot\vec{b_{2}})} \hspace{2mm} ,
\label{A2}
\end{equation}

\noindent
where $L_{1}= \vec{a_{1}}\cdot\vec{a_{2}} + \vec{b_{1}}\cdot\vec{b_{2}}$ and $L_{2}= \vec{a_{1}}\cdot\vec{b_{1}} + \vec{a_{1}}\cdot\vec{b_{2}} + \vec{a_{2}}\cdot\vec{b_{1}} + \vec{a_{2}}\cdot\vec{b_{2}}$ are the linear terms, $Q_{1}= (\vec{a_{1}}\cdot\vec{b_{1}})(\vec{a_{2}}\cdot\vec{b_{2}}) + (\vec{a_{1}}\cdot\vec{b_{2}})(\vec{a_{2}}\cdot\vec{b_{1}})$ and $Q_{2}= (\vec{a_{1}}\cdot\vec{a_{2}})(\vec{b_{1}}\cdot\vec{b_{2}})$ the quadratic terms and $x_{i} (i=1,..,5)$ are unknown constants to be determined. We can determine the constants using the following two facts:\\

\noindent
(a) If $\vec{a_{1}} = \vec{b_{1}}$ and $\vec{a_{2}} = \vec{b_{2}}$, the overlap must be 1. Using these in (\ref{A2}) reduces both the numerator and denominator to polynomials in $\vec{a_{1}}\cdot\vec{a_{2}}$, and the condition that the overlap be 1 can be satisfied only if the coefficients of the same powers of $\vec{a_{1}}\cdot\vec{a_{2}}$ in the numerator and the denominator are equal. This leads to the three equations

\begin{equation}
x_{1}+2x_{3}+x_{4} = 9 \hspace{2mm},\hspace{2mm} x_{2}+x_{3}=3 \hspace{2mm} \text{and} \hspace{3mm} x_{4}+x_{5}=1 \hspace{2mm} .
\label{A3}
\end{equation}

\noindent
(b) From the definition (\ref{A1}) one sees that the states $|\vec{a_{1}},\vec{a_{2}}\rangle$ and $|-\vec{a_{1}},-\vec{a_{1}}\rangle$ are orthogonal to each other. This implies that if  $\vec{b_{1}}=\vec{b_{2}}=-\vec{a_{1}}$ the overlap must be zero, but this can happen only if every power of $\vec{a_{1}}\cdot\vec{a_{2}}$ in the numerator vanishes and this leads to the pair of equations

\begin{equation}
x_{1}+x_{2}-2x_{3} = 0  \hspace{2mm} \text{and} \hspace{3mm} x_{2}-2x_{3}+2x_{4}+x_{5}=0 \hspace{2mm} .
\label{A4}
\end{equation}

\noindent
The solution to (\ref{A3}) and (\ref{A4}) is $x_{1}=3,x_{2}=1,x_{3}=2,x_{4}=2$ and $x_{5}=-1$, and putting these into (\ref{A2}) and doing some simplification leads to Eq.(\ref{olap2}) of the text.

\appendix
\section{\label{App2} Appendix 2: Proof of Proposition 6}

The proof of Proposition 6 involves finding the zeros of the function $F(\phi)+c_{4}$, as these will completely determine the other double-cone bases determined by the choice of $\big[D^{R}(\theta_{1},0)\big]$ as the initial basis. It is simplest to proceed by finding the zeros of the function $F(\phi^{\prime})+c_{4}$ and then to use the fact that $\phi = \phi^{\prime}+\phi_{0}$, with $\phi_{0}$ given by (\ref{ph02}). On using (\ref{Fprime2}),(\ref{d2p}),(\ref{c3p}) and (\ref{c4}), we see that the unbiasedness condition, $F(\phi^{\prime})+c_{4}=0$, assumes the simple form

\begin{equation}\label{unbcon}
 \sin^{2}\theta_{1}\sin^{2}\theta_{2}\cos2\phi^{\prime}+  d^{\prime}_{1}\cos\phi^{\prime} = 0
\end{equation}

\noindent
The expression on the left side of (\ref{unbcon}), regarded as a function of $\phi^{\prime}$, is the symmetrical function shown in Fig.4. If there is to be a basis on the double-cone $\theta_{2}$ that is unbiased to the basis $\big[D^{R}(\theta_{1},0)\big]$, then (\ref{unbcon}) must have three zeros that are spaced at an angle of $\pm2\pi/3$ from each other. This will happen only if the function in Fig.4 is positioned so that its smaller maximum just touches  the $x$-axis at the origin, for its other two zeros will then occur at $\pm2\pi/3$. From (\ref{unbcon}) we see that this happens when\\

\noindent
EITHER  \hspace{5mm} $ d^{\prime}_{1} = \sin^{2}\theta_{1}\sin^{2}\theta_{2}$   \hspace{5mm}and\hspace{5mm} $\phi^{\prime}=\frac{\pi}{3},\pi,\frac{5\pi}{3}$
\\
\noindent
OR  \hspace{10mm} $ d^{\prime}_{1} = -\sin^{2}\theta_{1}\sin^{2}\theta_{2}$   \hspace{3mm} and \hspace{5mm} $\phi^{\prime}=0,\frac{2\pi}{3},\frac{4\pi}{3}$\\

\noindent
These conditions can be combined into one and written as
\begin{equation}\label{unbcon2}
d^{\prime}_{1} =\sigma_{1} \sin^{2}\theta_{1}\sin^{2}\theta_{2} \hspace{2mm} \text{and} \hspace{2mm}
\phi^{\prime}=\frac{\pi}{6}(\sigma_{1}+1),\frac{\pi}{6}(\sigma_{1}+1)+\frac{2\pi}{3},\frac{\pi}{6}(\sigma_{1}+1)+\frac{4\pi}{3}
\end{equation}

\noindent
where $\sigma_{1}$ can be either $+1$ or $-1$. Squaring the first relation in (\ref{unbcon2}) and using (\ref{d1p}) transforms it into

\begin{multline}\label{unbcon3}
 \sin^{4}\theta_{1}\sin^{4}\theta_{2}=4\sin^{2}\theta_{1}\sin^{2}\theta_{2}\Big[\cos^{2}\theta_{1}\cos^{2}\theta_{2}  \big(1+\cos\phi_{1}\big)\big(1+\cos\phi_{2}\big) \\
 +\big(1-\cos\phi_{1}\big)\big(1-\cos\phi_{2}\big)+2\cos\theta_{1}\cos\theta_{2}\sin\phi_{1}\sin\phi_{2}\Big]
\end{multline}

\noindent
If one keeps the term involving $\sin\phi_{1}\sin\phi_{2}$ on one side of the equation and transposes the remaining terms to the other side, and then squares both sides, remembering that $\cos\phi_{i}=2\cot^{2}\theta_{i}$, with $\hspace{1mm} i=1\hspace{1mm} \text{or}\hspace{1mm}2$, one arrives at the quartic equation (\ref{quartic}), with $y=\cos^{2}\theta_{1}$ and the two real roots $x=\cos^{2}\theta_{2}$ and $x=\cos^{2}\theta_{3}$ determining the double-cones on which the other two bases of the MUB lie.\\

\noindent
To fix the bases on these double-cones, recall that the quartic was arrived at by squaring two equations: the first was in the transition from (\ref{unbcon2}) to (\ref{unbcon3}) and the second in the transition from (\ref{unbcon3}) to the quartic. To restore the lost information we should consider (\ref{unbcon2}) in its expanded form (\ref{offset}), with the variables $\sigma_{1}$ and $\sigma_{2}$ added, and see which combination of values allows the equation to be satisfied. This allows the bases on the double-cones to be fixed in the manner explained in the table below Eq.(\ref{offset}).

\appendix
\section{\label{App3} Appendix 3: Double-cone states of a MUB as rays in $\mathbb{CP}^{2}$}

First we recall \cite{majorana}-\cite{zimba} how the spin-1 state $|\vec{a_{1}},\vec{a_{2}}\rangle = (\theta_{1},\phi_{1}|\theta_{2},\phi_{2})$ can be expressed as a ray in $\mathbb{CP}^{2}$. If $\alpha_{1}=\tan(\theta_{1})e^{i\phi_{1}}$ and $\alpha_{2}=\tan(\theta_{2})e^{i\phi_{2}}$ are the complex parameters of the state, the ray corresponding to it is

\begin{equation}\label{A2.1}
      \hspace{10mm}  (1,\frac{\alpha_{1}+\alpha_{2}}{\sqrt{2}},\alpha_{1}\alpha_{2}) \hspace{2mm} .
\end{equation}

\noindent
If one of the complex parameters, say $\alpha_{2}$, is $\infty$ the ray is $(0,1,\sqrt{2}\alpha_{1})$ and if both parameters are $\infty$ the ray is $(0,0,1)$. \\

\noindent
 Consider the double-cone state $D^{R}(\theta_{a},0) = (\theta_{a},0|\pi-\theta_{a},\pi-\phi_{a})$, where $\phi_{a} = \cos^{-1}(2\cot^{2}\theta_{a})$. On working out its complex parameters and using (\ref{A2.1}), we see that the ray corresponding to this state is

\begin{equation}\label{A2.2}
  D^{R}(\theta_{a},0) = \Big(1, \frac{1}{\sqrt{2}}\big(\tan\frac{\theta_{a}}{2} - \cot\frac{\theta_{a}}{2} e^{-i\phi_{a}}\big),-e^{-i\phi_{a}}\Big)  \hspace{2mm} \rightarrow
  (1, e^{i\chi}, -e^{-i\phi_{a}}) \hspace{2mm} ,
\end{equation}

\noindent
where in the last step we wrote the middle component as $e^{i\chi}$ because a simple calculation shows it to have modulus unity. The time-reversed state of $D^{R}(\theta_{a},0)$, which is $D^{L}(\theta_{a},-\phi_{a})= (\theta_{a},-\phi_{a}|\pi-\theta_{a},\pi)$, can be obtained from (\ref{A2.1}) as

\begin{equation}\label{A2.3}
  D^{L}(\theta_{a},-\phi_{a}) = \Big(1, \frac{1}{\sqrt{2}}e^{-i\phi_{a}}\big(\tan\frac{\theta_{a}}{2}-\cot\frac{\theta_{a}}{2}e^{i\phi_{a}}\big),-e^{-i\phi_{a}}\Big)  \hspace{2mm}
  \rightarrow (1, e^{i\chi'}, -e^{-i\phi_{a}}) \hspace{2mm} ,
\end{equation}

\noindent
where again we have written the middle component as $e^{i\chi'}$ because it has modulus unity. On comparing (\ref{A2.2}) with(\ref{A2.3}) we see that $\chi' = -\phi_{a} -\chi$. The simplest way of fixing the values of $\chi$ and $\chi'$ is by using the fact that the states $D^{R}(\theta_{a},0)$ and $D^{L}(\theta_{a},-\phi_{a})$ are unbiased and that their (normalized) rays have an overlap of $1/3$. Then (\ref{A2.2}) and (\ref{A2.3}) imply that

\begin{equation}\label{A2.4}
  \cos(\chi-\chi') = -\frac{1}{2} \hspace{2mm}  \Rightarrow \hspace{2mm} \chi = \frac{\pi}{3}-\frac{\phi_{a}}{2} \hspace{2mm} \text{or}
 \hspace{2mm} \chi = \frac{2\pi}{3}-\frac{\phi_{a}}{2} \hspace{2mm} \hspace{2mm} .
\end{equation}

\noindent
However only the second of these possibilities satisfies (\ref{A2.2}) and we therefore find that $\chi = \frac{2\pi}{3}-\frac{\phi_{a}}{2}$ and
$\chi' = -\frac{2\pi}{3}-\frac{\phi_{a}}{2}$. Using these in (\ref{A2.2}) and (\ref{A2.3}) gives the form of these rays listed in Table \ref{tab1}.\\

\noindent
Next consider the state $D(\theta_{b},\pi - \phi_{a}/2)=(\theta_{b},\pi - \phi_{a}/2|\pi-\theta_{b},-\phi_{a}/2)$. From (\ref{A2.1}) and the expression for $\theta_{b}$ given in the caption to Table \ref{tab1} we can work out the ray corresponding to it as

\begin{equation}\label{A2.5}
  D(\theta_{b},\pi - \phi_{a}/2))  = (1, e^{-i\phi_{a}/2},-e^{-i\phi_{a}}) \hspace{2mm} ,
\end{equation}

\noindent
which agrees with the result given in Table \ref{tab1}.\\

\noindent
The other states in Table \ref{tab1} have M-vectors that are obtained by rotating the M-vectors of the states already considered by $2\pi/3$ or $4\pi/3$ about the z-axis, and the corresponding rays can be written down by multiplying the second and third components of the original rays by $\omega$ and $\omega^{2}$ (for the states rotated by $2\pi/3$) or by $\omega^{2}$ and $\omega$ (for the states rotated by $4\pi/3$).


\end{document}